\documentclass[journal]{IEEEtran}

\usepackage{multirow}
\usepackage{subcaption}
\usepackage{makecell}
\usepackage{tabularx}
\usepackage{cite}
\usepackage{array}
\usepackage{caption}
\usepackage{amsmath}
\usepackage{amsfonts}
\usepackage{pdfpages}
\usepackage{graphicx}
\usepackage{footnote}
\usepackage{booktabs}
\usepackage{diagbox}
\usepackage{arydshln}
\usepackage{xcolor}
\usepackage{pgfplots}

\usepackage[utf8]{inputenc}
\usepackage[english]{babel}
\usepackage{fancyhdr}

\pgfplotsset{compat=1.7}
\usetikzlibrary{patterns}

\usepackage[pdfencoding=auto]{hyperref}
\hypersetup{
    colorlinks=true,
    linkcolor=blue,
    filecolor=blue, 
    urlcolor=black,
    citecolor=blue,
}

\begin{document}


\title{Unraveling Attacks in Machine Learning-based IoT Ecosystems: A Survey and the Open Libraries Behind Them}
\author{Chao~Liu,~\IEEEmembership{Member,~IEEE,}
        Boxi Chen*,~\IEEEmembership{Student Member,~IEEE,}
        Wei Shao,~\IEEEmembership{Member,~IEEE,}
        Chris Zhang,~\IEEEmembership{Senior Member,~IEEE,}
        Kelvin K.L. Wong,~\IEEEmembership{Member,~IEEE,}
        and~Yi Zhang,~\IEEEmembership{Fellow,~IEEE}% <-this % stops a space

\thanks{Chao Liu and Boxi Chen* contributed equally to this work. This work was completed during Boxi Chen's internship.}
\thanks{Corresponding author: Kelvin K.L. Wong (kelvin.wong@usask.edu) and Yi Zhang (zhangyi@scu.edu.cn)}
%Boxi Chen (bxchen824@gmail.com)
\thanks{Chao Liu was with the Department of Electrical and Computer Engineering, University of Maryland, Baltimore County, MD, USA.}% <-this % stops a space
\thanks{Boxi Chen was with the Department of Research, Deep Red Future Technology Co. Ltd., Shenzhen, China.}% <-this % stops a space
\thanks{Wei Shao is with the data61, CSIRO, Australia.}
\thanks{Chris Zhang and Kelvin K.L. Wong were with the Department of Mechanical Engineering, University of Saskatchewan, Saskatchewan, Canada.}% <-this % stops a space
\thanks{Yi Zhang was with Machine Intelligence Laboratory, College of Computer Science, Sichuan University, Sichuan, China.}% <-this % stops a space
}

\markboth{IEEE INTERNET OF THINGS JOURNAL, October~2023}%
{Shell \MakeLowercase{\textit{et al.}}: Bare Demo of IEEEtran.cls for IEEE Journals}

\maketitle
\thispagestyle{empty} 

\begin{abstract}
The advent of the Internet of Things (IoT) has brought forth an era of unprecedented connectivity, with an estimated 80 billion smart devices expected to be in operation by the end of 2025. These devices facilitate a multitude of smart applications, enhancing the quality of life and efficiency across various domains. Machine Learning (ML) serves as a crucial technology, not only for analyzing IoT-generated data but also for diverse applications within the IoT ecosystem. For instance, ML finds utility in IoT device recognition, anomaly detection, and even in uncovering malicious activities. This paper embarks on a comprehensive exploration of the security threats arising from ML's integration into various facets of IoT, spanning various attack types including membership inference, adversarial evasion, reconstruction, property inference, model extraction, and poisoning attacks. Unlike previous studies, our work offers a holistic perspective, categorizing threats based on criteria such as adversary models, attack targets, and key security attributes (confidentiality, availability, and integrity). We delve into the underlying techniques of ML attacks in IoT environment, providing a critical evaluation of their mechanisms and impacts.
Furthermore, our research thoroughly assesses 65 libraries, both author-contributed and third-party, evaluating their role in safeguarding model and data privacy. We emphasize the availability and usability of these libraries, aiming to arm the community with the necessary tools to bolster their defenses against the evolving threat landscape. Through our comprehensive review and analysis, this paper seeks to contribute to the ongoing discourse on ML-based IoT security, offering valuable insights and practical solutions to secure ML models and data in the rapidly expanding field of artificial intelligence in IoT.
\end{abstract}

\begin{IEEEkeywords}
Internet of Things, Machine Learning, Attack, Security, Open Library,  Artificial Intelligence
\end{IEEEkeywords}

\IEEEpeerreviewmaketitle

\section{\textbf{Introduction}}
\subsection{\textbf{IoT and Machine Learning}}

\IEEEPARstart{T}{he} original design concept of the Internet of Things is centered on achieving connectivity among all things, aiming to create a digitalized and intelligent network system. Within this system, devices can communicate and collaborate, thereby simplifying and accelerating a variety of tasks and processes. In the IoT ecosystem, end-users can interact with IoT devices through their smartphones and computers \cite{neto2016aot}, which offers immense convenience in sectors like industry, household management, and healthcare. According to an IoT industry white paper jointly published by iResearch Consulting Group and Kingsoft Cloud, it is predicted \cite{rabah2018convergence} that by 2025, the global number of IoT connections will increase to 24.6 billion, and the global IoT revenue will grow to 1.1 trillion USD.

\begin{figure*}[!ht]   
     \centering
     \includegraphics[width=0.85\linewidth]{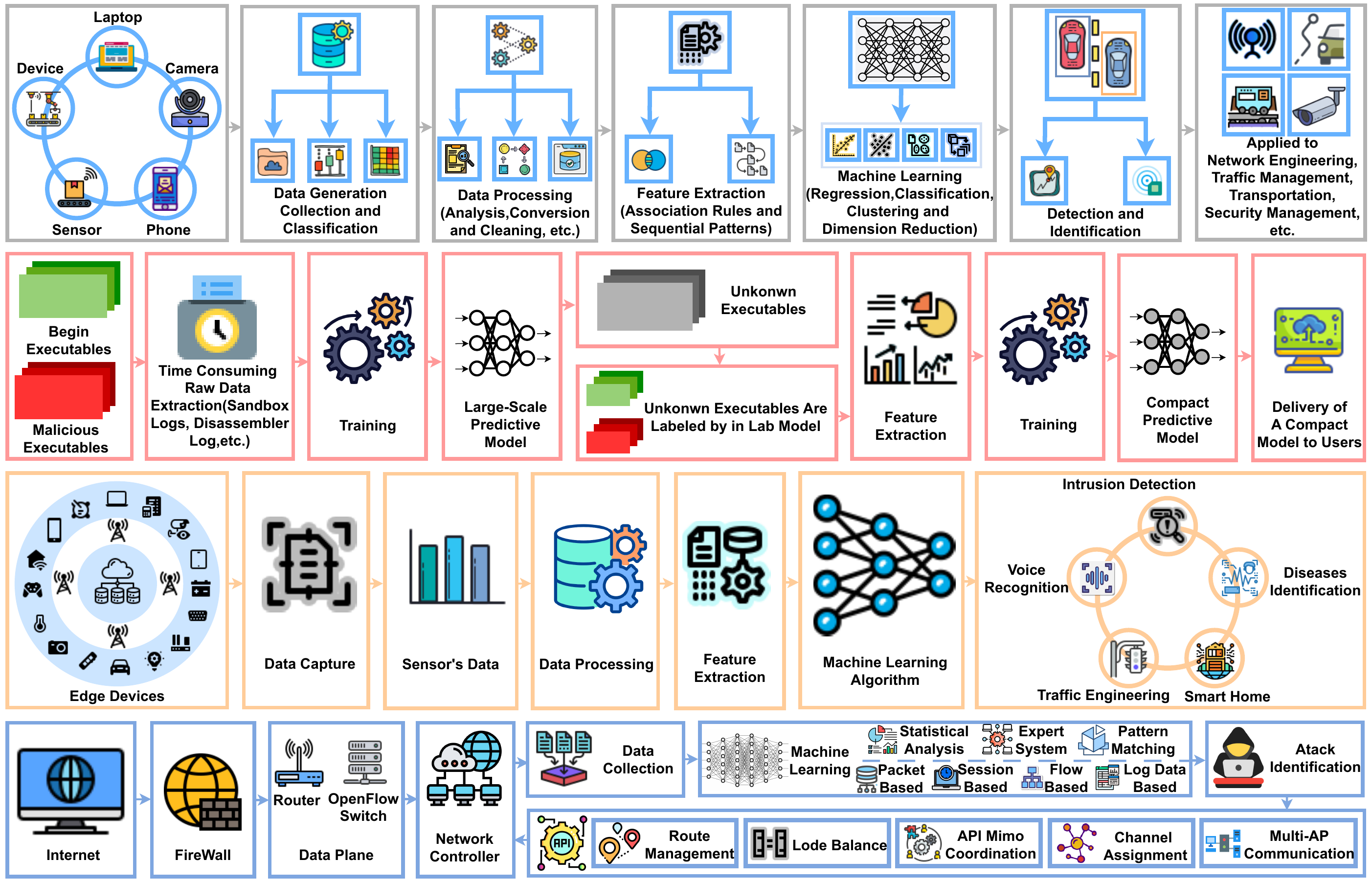}
     \caption{\textbf{Typical scenarios and ML-based IoT visions.}}
     \label{overviewfig1}
\end{figure*}

The foundational idea of IoT is to connect everything in the world to the internet, allowing for autonomous recognition, communication, and even decision-making among objects. The tremendous advancements in computer communication and related technologies have made numerous IoT applications possible. Researchers from diverse industries, research groups, academia, and government bodies are paying close attention to the IoT revolution, striving to create a more convenient environment comprised of an array of intelligent systems such as smart homes, intelligent transportation, global supply chain logistics, and healthcare \cite{atzori2010internet,miorandi2012internet, bandyopadhyay2011internet}. IoT’s adoption of automatic data sensing and collection has generated big data, a category of data that has surpassed the capacity of traditional manual data analysis and processing methods. This paradigm shift underscores the need for more advanced and automated data analysis solutions to harness the full potential of IoT’s capabilities.
One of the most effective methods to enhance the intelligence of data analysis and processing in the Internet of Things (IoT) is to integrate Machine Learning (ML) into its operations. ML systems shown in Fig. \ref{overviewfig1} are designed to automatically learn models from training data and utilize these models to make predictions. They find extensive applications across numerous fields, including ML-based IoT device identification, ML-based IoT Malware, etc.
\subsubsection{\textbf{ML-based IoT Device Identification}} Mirai \cite{antonakakis2017understanding}, in particular, stands as a large-scale botnet composed of compromised IoT devices. Its massive Distributed Denial-of-Service (DDoS) attack in 2016 paralyzed crucial communication infrastructures along the U.S. East Coast, resulting in substantial economic losses. 

Research on IoT device identification and anomaly detection is crucial for asset and security management. These two areas are closely interconnected, sharing similarities on a technical level, as both can be viewed as classification problems. The process involves data collection: IoT device identification collects passive network traffic or response data packets, while anomaly detection gathers normal traffic data or data about abnormal patterns, logs, or application source codes. After data collection, preprocessing occurs. Device identification extracts rules for a rule repository, and anomaly detection constructs a normative behavior profile, identifying deviations as anomalies. Both methods leverage preprocessed data to derive raw features, building machine learning models. Device identification yields device categories, while anomaly detection classifies as normal/abnormal or specific abnormal types. Evaluation involves metrics like Accuracy, Precision, Recall, and F1 scores. Anomaly detection uses additional metrics like false alarm rate and false negative rate. Machine learning-based methods extract features to form device fingerprints. Subsequently, through machine learning-based techniques \cite{yang2019towards,caballero2007fig}, classification models are trained using labeled data, and these models are employed for device identification. Therefore, the selection and extraction of features, methods for data labeling, and the choice and training of models constitute focal points in research.
\subsubsection{\textbf{ML-based IoT Malware Detection}}In recent years, security researchers have shifted focus to detecting malicious software in the Internet of Things (IoT). Traditional methods relied on feature databases and manual analysis, but the exponential growth of malicious software created inefficiencies and challenges in addressing unknown security risks \cite{griffioen2020examining, ccetin2019cleaning}. To overcome these issues, researchers are exploring the application of artificial intelligence (AI) techniques \cite{downing2021deepreflect, alrawi2021forecasting, wang2020you}. While AI-based research in IoT malicious code detection is gaining traction, challenges unique to IoT devices need attention \cite{jamal2022malware}: Diverse Architectures: Malicious code in IoT can infect devices with various CPU architectures, making it challenging to directly apply matured detection methods from conventional devices. Resource Limitations: IoT devices, having smaller form factors and resource constraints, require lightweight detection systems due to limited memory space and low power. Addressing these challenges is crucial for the effectiveness of AI-based detection in IoT malicious code while enhancing detection accuracy.
\subsubsection{\textbf{Edge Computing with ML}}
ML applications often face computational challenges, especially on resource-constrained local devices in industrial IoT. While cloud computing offers a solution by offloading tasks to servers, the drawback is significant response latency due to the distance between devices and servers. Edge computing emerges as a promising solution, deploying servers near IoT devices for faster task processing and reduced network loads \cite{qiu2020edge}. Different computational capabilities among IoT devices, edge servers, and cloud servers allow deployment of varying complexity ML models. However, executing ML tasks on these platforms results in varying inference precision. Lower precision impacts system performance, potentially causing material waste or delays. Despite this, existing optimizations often overlook the impact of ML model complexity on precision. Balancing execution speed, cost reduction, and ensuring precision is crucial for industrial IoT safety and efficiency. Limited storage on edge servers poses challenges in deploying diverse ML models, while computational resource constraints compromise task processing efficiency. Addressing these challenges is essential for optimizing system performance in industrial IoT scenarios \cite{kato2020ten}.
\subsubsection{\textbf{Software‑defined Networking with ML}} Software-defined networking (SDN) has recently gained traction in both academic and industrial sectors due to its inherent flexibility. By decoupling the control plane from the forwarding plane, SDN empowers network operators to manipulate networks using high-level configuration languages, alleviating the need for intricate forwarding table configurations. In the context of IoT, where device complexity and diversity are pronounced, configuring the data path in IoT networks becomes notably challenging compared to traditional networks. Therefore, SDN emerges as a pivotal solution in the IoT landscape \cite{kim2013improving}. 

However, owing to the intricate nature of IoT systems, leveraging machine learning within the control plane becomes imperative for more effective network management. The complexities inherent in IoT necessitate adaptive and intelligent network control mechanisms, which machine learning can facilitate. ML has two important aspects for SDN in IoT network: (1) ML facilitates IoT network management \cite{kim2017identifying, vukobratovic2016condense,jagadeesan2016programming} with SDN, enhancing ease and efficacy.
(2) ML aids in detecting potential intrusions \cite{
uwagbole2017applied,ahmed2017mitigating,bhunia2017dynamic}, thereby improving accuracy.
\subsubsection{\textbf{ML in IoT applications}}IoT applications, prevalent in healthcare, agriculture, and industrial sectors, face challenges due to diverse and complex data sources. The model of IoT applications includes wearable devices, mobile sensors, network cameras, and various wireless sensors capturing human vital signs and environmental data. Machine learning methods process this data for tasks like health monitoring, activity recognition, fraud detection, and object identification.

Studies focus on personal health monitoring and industrial use cases. IoT networks excel in stress monitoring, human activity recognition, and disease prediction in agriculture. Madeira et al. \cite{madeira2016machine} introduced a system predicting human presence using diverse interaction data from IoT devices. Siryani et al. \cite{siryani2017machine} applied machine learning to optimize smart meter operations, reducing costs by predicting technician dispatch necessity. Instances like \cite{walinjkar2017ecg, nguyen2017review, ling2017identifying, guo2015automated} showcase machine learning applications in IoT.

In summary, machine learning models trained on data from IoT devices have significantly enhanced efficiency across industries like healthcare, logistics, and urban planning. However, these devices collect sensitive information, posing privacy and security risks. Adversaries could misuse data for unfair practices, impacting individuals and even national security. For instance, smart fitness tracker data could lead to unfair health insurance adjustments, while compromising IoT devices in critical infrastructure may result in disruptions and security threats. Despite the benefits, the privacy and security concerns surrounding sensitive training data have become crucial, drawing attention from academia and industry stakeholders.

\subsection{\textbf{IoT-integrated Machine Learning Security Risks}}
\begin{figure}[!ht]   
     \centering
     \includegraphics[width=0.8\linewidth]{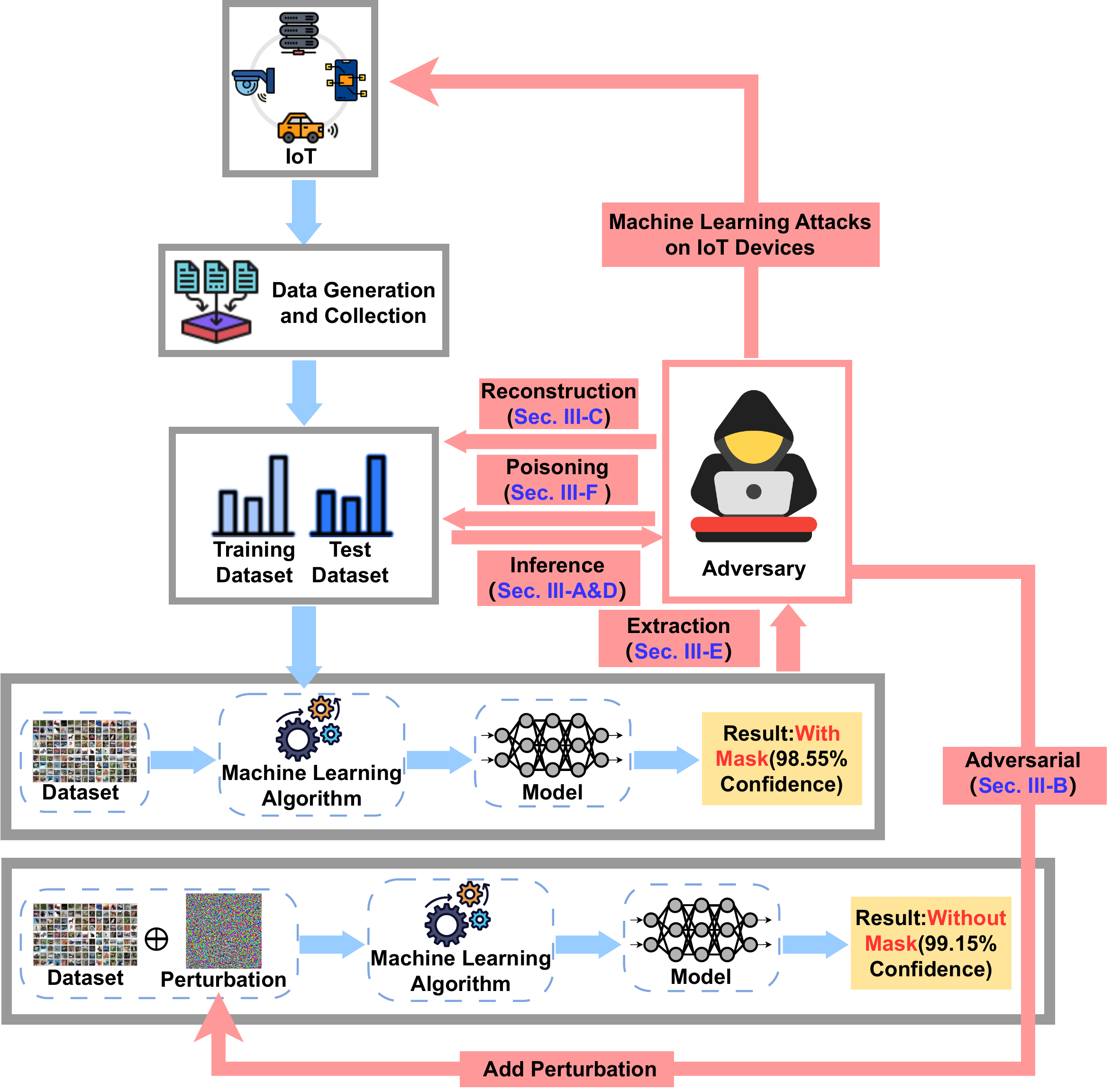}
     \caption{\textbf{Different attack models targeting ML-based IoT.}}
     \label{attackonIOT}
\end{figure}
The rapid development of AI and IoT technologies has brought substantial progress and innovation, offering diverse capabilities and services across domains. However, this progress comes with challenges, particularly in the realms of security risks and vulnerabilities. In the interconnected world of IoT devices and systems, these challenges extend to data privacy, network security, and ethical concerns. This article explores six prevalent attack types on machine learning systems in IoT Fig. \ref{attackonIOT}: membership inference, adversarial attacks, reconstruction attacks, property inference attacks, model extraction attacks, and poisoning attacks. Real-world instances highlight how these attacks can compromise the privacy and security of machine learning systems across various domains within the IoT environment. The intricately connected ecosystem of IoT devices, from smart thermostats to industrial sensors, creates a vast attack surface, making ML systems susceptible to malicious activities. The insights provided aim to illuminate these vulnerabilities, offering a comprehensive overview of potential threats and their implications on the privacy and security of machine learning systems within IoT networks and applications.
\subsubsection{\textbf{Membership Inference Attack Against Collaborative Inference in Industrial IoT}}
Despite the notable achievements across a spectrum of industrial IIoT applications, collaborative inference systems inherently face diverse privacy and security risks, including model inversion attacks and property inference attacks. Researchers \cite{chen2020practical} have developed practical membership inference attacks (MIA) targeting collaborative inference methods, aiming to ascertain whether an individual input sample exists within the training dataset.
\subsubsection{\textbf{Adversarial Attacks on Graph-based IoT
Malware Detection Systems}}
Machine learning algorithms, particularly deep learning networks, are actively involved in identifying and categorizing malicious software from benign ones. Yet, the increased reliance on deep learning models within security realms presents opportunities for adversaries to manipulate these models, steering them toward generating specific outputs of their choice. Research \cite{abusnaina2019adversarial} has unveiled vulnerabilities inherent in machine and deep learning networks. For instance, adversaries can coerce models into generating desired outputs, such as misclassifications, by creating adversarial examples (AEs).
\subsubsection{\textbf{Reconstruction Attack for ML-based Recognition Systems}}
 Given the extensive use of recognition systems in cloud environments, the risk of cyber threats remains significantly high. Consequently, cybersecurity measures have garnered substantial focus, leading to the proposal of diverse general-purpose algorithms aimed at combating cybercrimes across industrial control systems, power quality management, unmanned aerial vehicles (UAVs), smart grids, and other sectors. The model inversion attack (MIA) \cite{khosravy2022social} is a technique used to create a replicated data point associated with a specific user identity within recognition systems. This attack traces the impact of a corresponding user's training data on the structure and parameters of the system's model. The clone data generated through MIA often exhibit high-recognition scores and are erroneously accepted by deep-learning-based systems, despite the complexity of the model structure, numerous class labels, and extensive training data. Surprisingly, these factors, which might typically enhance security, fail to prevent the tracing of user features through the effect of their training data on the model structure.
 \subsubsection{\textbf{Property Inference Attack on Edge-Cloud Collaborative Inference Systems}}
With the substantial improvement in end device computation capabilities, collaborative learning, also known as federated learning, has evolved significantly within real-world IoT applications. An example of an inference attack, property inference attack \cite{ma2021nosnoop}, within the training process is the property (or attribute) inference attack, which aims to deduce whether a training batch contains data associated with a specific sensitive property, such as race, or not.
 \subsubsection{\textbf{AI Model
Extraction Attack in IoT
}}
In the realm of intelligent IoT security, safeguarding ML-based IoT models is of paramount importance. When these systems operate on hardware platforms (smart-X devices, such as automotive and virtual reality/augmented reality devices), inadvertent physical information leaks occur, such as energy consumption, execution time, and electromagnetic emissions released during data computation. Adversaries monitor these physical leaks in real-time through predefined probes and analyze them to infer sensitive information. AI models encompass various elements like structure, metaparameters, and weights, hence the extraction of ML-based IoT models involves multiple methods. By exerting physical control over the hardware, adversarys leverage multidimensional side-channel signals to extract the structure and parameters of AI models operating on the hardware. Memory access patterns and timing serve as common side-channel signals used to deduce the architecture of AI models \cite{pan2022side}.
\subsubsection{\textbf{Poisoning Attacks on Machine Learning in IoT Environments}}
Utilizing machine learning in IoT settings introduces distinct security challenges as adversaries can potentially tamper with sensors, manipulating the training data. This form of attack termed a poisoning attack, enables adversaries to significantly impair overall performance, induce targeted misclassifications or undesirable behavior, and implant backdoors or neural trojans \cite{liu2017neural}. A well-known instance of a poisoning attack beyond the realm of IoT transpired when Microsoft introduced Tay, a chat bot, designed to learn and generate human-like tweets. Users exploited the system by posting offensive content, leading Tay to produce similarly offensive tweets. Microsoft had to swiftly remove the bot within just 16 hours due to this manipulation.

In summary, ML attacks pose significant threats to IoT due to their potential to compromise security and privacy. These attacks, including model inversion attacks, poisoning attacks, adversarial attacks, etc, exploit vulnerabilities in IoT devices and systems. They can manipulate training data, compromise AI models, and extract sensitive information, risking the integrity and confidentiality of data collected by IoT devices. This compromises user privacy, creates avenues for unauthorized access, and undermines the reliability of IoT-driven applications, amplifying concerns about security and privacy in connected environments.
\begin{figure*}[!ht]   
     \centering
     \includegraphics[width=0.8\linewidth]{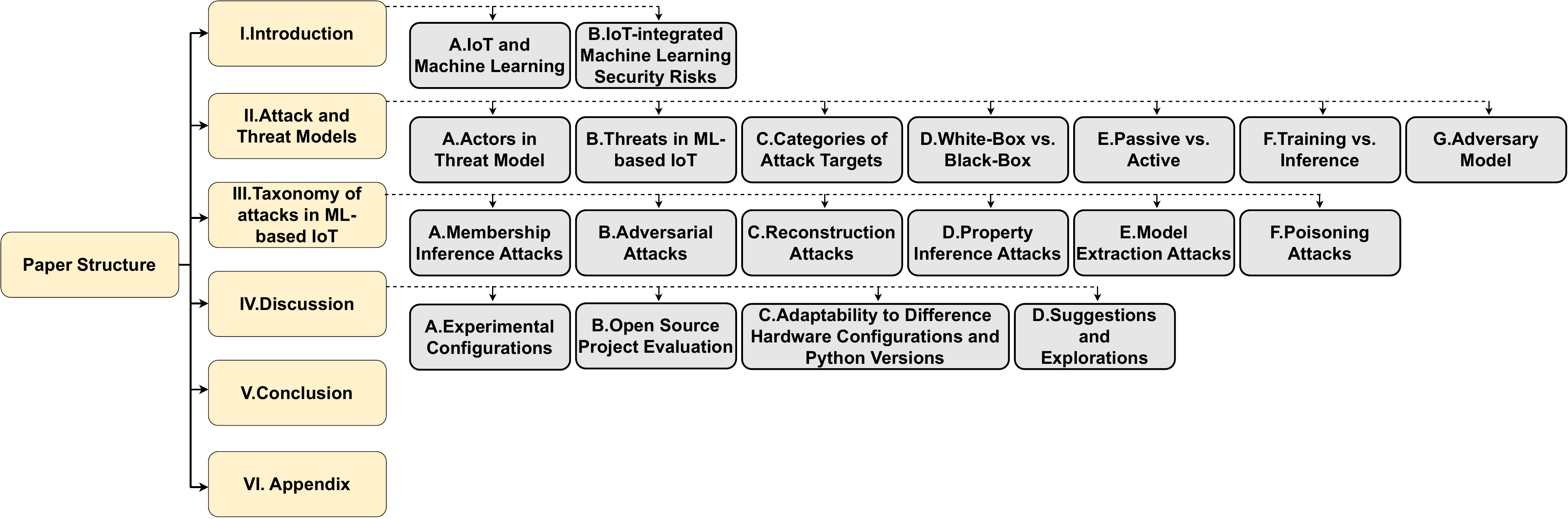}
     \caption{\textbf{The structure of this article.}}
     \label{Fig44}
\end{figure*}
The main contributions of this paper are as follows: 
\begin{itemize}
\item We delve into a comprehensive range of attack types pertinent to ML-integrated IoT systems, including membership inference attacks, adversarial attacks (evasion attacks), reconstruction attacks (such as model inversion and attribute inference attacks), property inference attacks, model extraction attacks, and poisoning attacks (covering both data and model poisoning). Unlike other studies that might focus narrowly on specific security attack facets, our investigation spans the entire spectrum of security threats, emphasizing their implications in IoT environments.
\item Our categorization of machine learning attacks takes into account the multifaceted nature and objectives of these threats in IoT settings. For example, we differentiate between white-box and black-box attacks based on the attacker’s level of knowledge. These attacks are further classified into data-centric and model-centric types. In terms of security, we organize them under the umbrellas of confidentiality, availability, and integrity threats. Furthermore, we provide a concise summary of key technologies discussed in our paper, aiding in a more structured and nuanced understanding of the security challenges in ML-enhanced IoT systems.
\item In our exploration of ML security within IoT settings, we've assessed 65 libraries. These libraries specifically focus on addressing model and data privacy within machine learning applications relevant to IoT. They encompass offerings developed by authors as well as those provided by various individuals or entities. Our evaluation spans diverse facets and objectives of ML systems within IoT scenarios, introducing a comprehensive set of metrics for assessment. We aim to verify the accessibility, usability, and practical applicability of these libraries in securing IoT systems. Through rigorous testing and validation, we're advocating for the usability of the majority of these libraries in real-world IoT applications.
\end{itemize}

The rest of this paper is organized as follows: Section II describes the attack and threat models. Section III describes the taxonomy of attacks in ML-based IoT. Section IV focuses on the discussion of this paper. At last, in Section V, we conclude the paper. The roadmap and organization of this article are shown
in Fig. \ref{Fig44}.

\section{\textbf{Attack and Threat Models}}
Understanding and defending against privacy-related attacks in ML-based IoT requires a comprehensive model of the environment, the involved actors, and the assets requiring protection. In the context of a threat model, the sensitive assets susceptible to potential attacks encompass the training dataset $\mathcal{D}$, the model, its parameters $\sigma$, hyper-parameters, and its architecture.
\begin{figure*}[!ht]   
     \centering
     \includegraphics[width=0.8\linewidth]{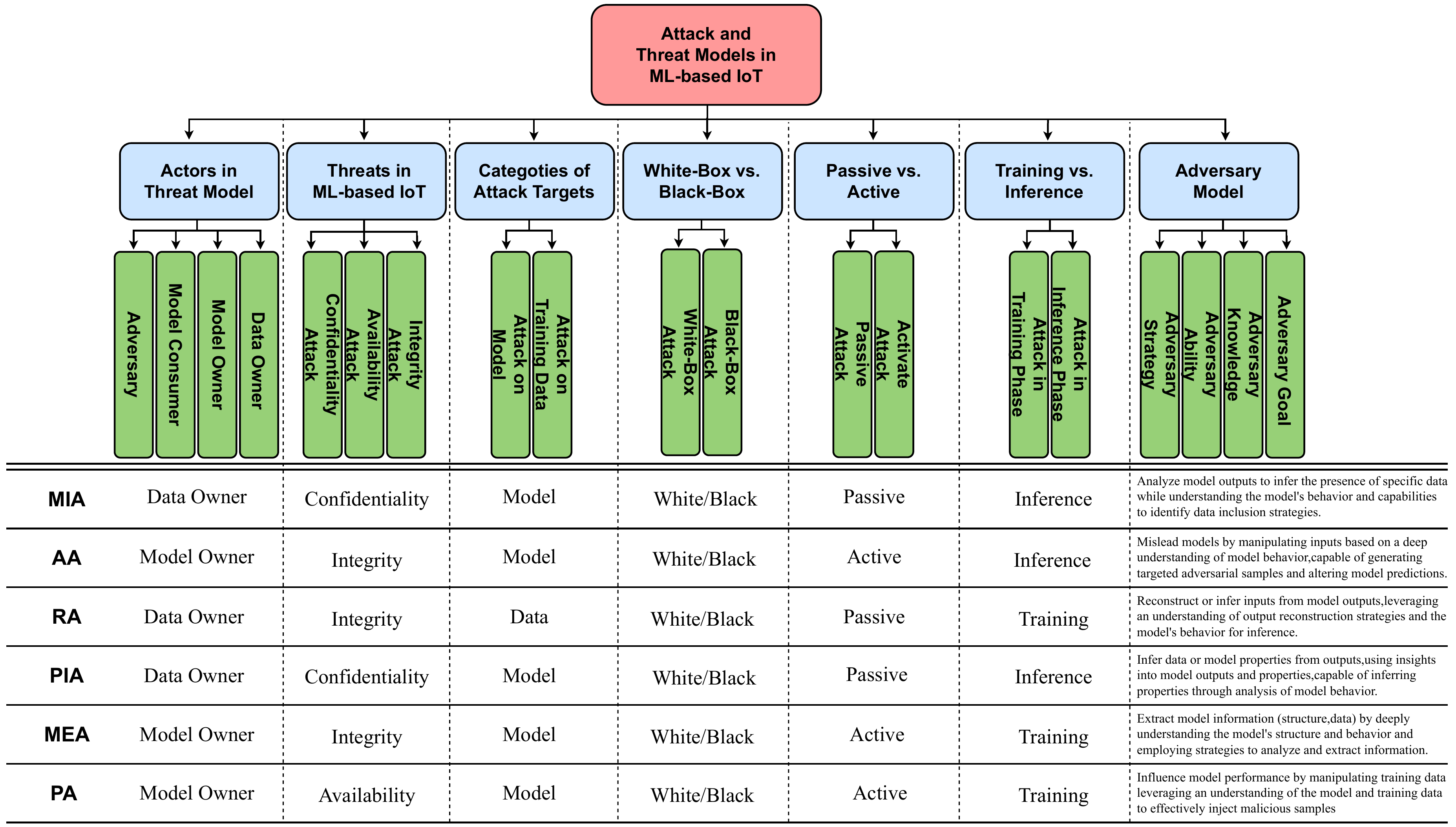}
     \caption{\textbf{Different categories in attack and threat models.}}
     \label{Fig}
\end{figure*}
\subsection{\textbf{Actors in Threat Model}}
\subsubsection{\textbf{Data owners}} are individuals or entities whose data holds sensitivity or confidentiality.
\subsubsection{\textbf{Model owners}} are individuals or entities who possess the machine learning models. They may or may not be the owners of the underlying data and might have varying preferences regarding sharing information about their models.
\subsubsection{\textbf{Model consumers}} are individuals or entities utilizing the services provided by the model owner, typically accessed through programming interfaces or user interfaces.
\subsubsection{\textbf{Adversaries}} refer to individuals or entities that, similar to normal consumers, might access the model's interfaces. In certain cases, if permitted by the model owner, adversaries may gain access to the model itself.

\subsection{\textbf{Threats in ML-based IoT}}
Threats in ML-based IoT can be divided into two categories:  security attacks and privacy infringements. In the former, attackers compromise the integrity and availability of data, while in the latter, attackers violate the privacy and confidentiality of data. Common attacks include the following three types.
\subsubsection{\textbf{Integrity Attacks}}
Integrity attacks \cite{abosata2021internet} aim to alter the behavior or output of a machine learning system by maliciously manipulating training data or models. For example, by injecting fictitious data into the training data set, it may lead to inaccurate classification and erode user trust in the system. This is similar to deliberately mixing substandard products with high-quality products during quality inspections, thereby reducing the credibility of the entire product. In an adversarial attack, the input data is modified to fool the model, leading to incorrect classification. In the Internet of Things, tampering with sensor data for predictive maintenance may mislead the model, leading to incorrect predictions or inappropriate maintenance operations, thus affecting the functionality and reliability of the equipment.
\subsubsection{\textbf{Availability Attacks}}
Availability attacks \cite{xu2003sustaining} aim to disrupt the normal functionality of machine learning-based IoT or generate inaccurate output, leading to system crashes, service interruptions, or erroneous results. This is similar to blocking traffic causing congestion or interfering with communications to disrupt messaging. In IoT, denial-of-service attacks overwhelm a smart home automation system with too many commands, rendering it unable to respond to legitimate requests and disrupting the functionality of the entire network. Likewise, flooding sensor networks with too much or erroneous data can lead to delays or system failures in making timely decisions based on sensor inputs.
\subsubsection{\textbf{Confidentiality Attacks}}
Confidentiality attacks \cite{nawir2016internet} aim to acquire sensitive information or private data within ML systems, analogous to a thief stealing from a secure vault or a hacker illicitly taking user's personally identifiable information. In IoT, these attacks may lead to unauthorized access, theft, or leakage of users' sensitive data, posing threats to privacy, trade secrets, or national security. Attackers can obtain training data or model parameters through various means, like side-channel attacks where monitoring power consumption patterns reveals details about processed inputs. Using model inversion techniques \cite{wang2023nvleak}, an attacker may reconstruct facial features from a facial recognition system's output.

\subsection{\textbf{Categories of Attack Targets}}
\subsubsection{\textbf{Attack on Training Data}} Protecting the privacy of training data \cite{kosut2010malicious} is crucial, especially in IoT scenarios. For example, a company creating a predictive model based on data from IoT sensors in smart cities aims to optimize traffic flow without compromising individual privacy. Ensuring training data privacy involves: a) Data Confidentiality: Concealing specific IoT device data or individual movements to prevent exposure of identifiable information; b) Feature Anonymization: Protecting sensitive attributes within IoT data to preserve individual or entity privacy; c) Securing Statistical Insights: Ensuring statistical patterns derived from IoT data do not compromise privacy; d) Protecting Membership Information: Concealing whether specific IoT devices or data points were part of the training set to prevent inference about their inclusion.
\subsubsection{\textbf{Attack on Model}} Privacy concerns \cite{luo2020adversarial} in machine learning extend beyond data to model components, including parameters and algorithms. For example, an IoT-based company developing a model for forecasting energy consumption in smart homes faces risks to its commercial value and intellectual property if model specifics are exposed. Similarly, commercial IoT-based machine learning services, relying on proprietary models and algorithms, could experience substantial revenue loss if these details are revealed. In summary, attacks on machine learning systems in IoT can target not just data but also the structure, parameters, or algorithms, posing risks to intellectual property, commercial viability, and revenue streams.
\subsection{\textbf{White-Box vs. Black-Box}}
In the context of ML-based IoT attacks, attack methods can be classified as either white-box attacks or black-box attacks depending on the level of adversarial knowledge.
\subsubsection{\textbf{White-Box Attacks}} A white-box attack \cite{biryukov2018attacks} occurs when the attacker possesses detailed knowledge of the target model, including the model’s internal structure, parameters, weights, and training data. An attacker may have access to all information about the model.
\subsubsection{\textbf{Black-Box Attacks}} A black-box attack\cite{bhagoji2018practical} means that the attacker can only access the input and output of the target model but lacks knowledge about the model's internal structure, parameters, and weights. The attacker does not have access to the details of the target model. Therefore, the adversary can infer information about the model by carefully designing inputs and observing the model's outputs.
Black-box attacks are generally more challenging because the attacker needs to reverse engineer the behavior of the target model through experimentation and observation without direct visibility into its internals.

\subsection{\textbf{Passive vs. Active}}
 “Active attacks” and “passive attacks” are two terms used to describe the goals and strategies of attackers. These two types of attacks have different characteristics and objectives, and understanding them helps in better defending against potential threats.
\subsubsection{\textbf{Active Attacks}} An active attack\cite{sampigethaya2006survey} occurs when an attacker takes deliberate actions to disrupt, corrupt, or modify ML-based IoT models or training data to achieve their malicious objectives.
\subsubsection{\textbf{Passive Attacks}} A passive attack\cite{serjantov2003passive} happens when an attacker observes, listens, or collects information without altering or interfering with the ML-based IoT model or training data. In passive aggression, the attacker's primary aim is to acquire sensitive information or comprehend the model's behavior without causing noticeable interference.

\subsection{\textbf{Training vs. Inference}}
\subsubsection{\textbf{Attack in Training Phase}}The adversary aims to gain insights into the model, potentially by accessing summaries, partial segments, or the entirety of the training data. They might endeavor to construct a substitute model, also known as an auxiliary model, which they can then utilize to launch attacks on the victim system.
\subsubsection{\textbf{Attack in Inference Phase}}The adversary gathers information about the model's characteristics by observing the inferences it generates.
\subsection{\textbf{Adversary Model}}
Adversary models \cite{bhardwaj2023adversarial} are theoretical frameworks used in security research and practice to describe and analyze the capabilities, intentions, and behaviors of potential attackers. In ML-based IoT security, adversary models are often employed to characterize the strength of an adversary. Barreno et al. \cite{barreno2010security} in 2010 considered adversary models that account for the capabilities of attackers and their objectives. Building upon the work of Biggio et al. \cite{biggio2013security} in 2013 further refined these models, proposing adversary models encompassing adversary objectives, knowledge, capabilities, and strategies. Current research widely accepts this framework as the standard adversary or attacker model. By delineating adversaries along these four dimensions, it becomes possible to systematically assess the extent of the threat posed by the adversary.
\subsubsection{\textbf{Adversary Goal}}
The adversary's target, as discussed by \cite{biggio2013security}, involves both the extent of destruction and the level of specificity they aim for. Destruction includes integrity, availability, and privacy, while specificity encompasses targeted and non-targeted objectives. Destruction of integrity involves unauthorized actions on data, like addition, deletion, modification, or destruction, impacting the accuracy of trained models, for example, in medical data. Destruction of availability aims to make the target service unusable, achieved by injecting malicious data into the training dataset, rendering the model ineffective. Destruction of privacy targets the theft of private data, extracting information from the training dataset. Regarding specificity, targeted objectives involve selectively disrupting specific aspects, like stealing the private information of a particular customer in medical data. Non-targeted objectives cause non-selective disruptions, impacting data integrity broadly.

\subsubsection{\textbf{Adversary Knowledge}}
Adversarial knowledge \cite{song2018machine} refers to the amount of information possessed by an adversary about the target model or the target environment, which includes details about the model's training data, model structure, parameters, information obtained through the model's outputs, and so on. Adversarial knowledge can be categorized based on the quantity of information the adversary possesses, namely limited knowledge or complete knowledge.

\subsubsection{\textbf{Adversary Ability}}
Adversary ability refers to the degree of control that an adversary has over both training and testing data. The impact of adversaries on data can be categorized as inductive, which affects the dataset, or exploratory, which does not affect the dataset. Adversary ability \cite{vsrndic2014practical} can also be defined based on whether the adversary can intervene in model training, access training datasets, collect intermediate results, etc. Depending on the level of control over data and models, adversaries can be further classified as strong or weak adversaries. Strong adversaries can intervene to a certain extent in model training, access training datasets, and collect intermediate results. In contrast, weak adversaries can only obtain information about the model or training data through attack methods.

\subsubsection{\textbf{Adversary Strategy}}
Adversary strategy pertains to the attack methods employed by adversaries, considering their objectives, knowledge, and capabilities to achieve optimal outcomes against the target. These adversary strategies typically involve utilizing different attack techniques at various stages of ML-based IoT. For instance, approaches like poisoning attacks are frequently employed during the training phase, whereas adversarial attacks and privacy attacks are commonly deployed during the prediction phase.

\begin{table}[ht!]
\centering
\resizebox{0.5\textwidth}{!}{
\begin{tabular}{ccccccc}
\hline
\diagbox{\textbf{Type}}{\textbf{Method}}& \textbf{Learning} & \textbf{GAN} & \textbf{Gradient-based} & \textbf{Shadow Training} & \textbf{Threshold} & \textbf{Equation-solving}\\ \hline
\textbf{MIA} & $\bullet$ & $\bullet$ & & $\bullet$ & $\bullet$ & \\ 
\textbf{AA} & $\bullet$ & $\bullet$ & $\bullet$ & & $\bullet$ & $\bullet$ \\ 
\textbf{RA} & $\bullet$ & $\bullet$ & $\bullet$ & & $\bullet$ & $\bullet$ \\ 
\textbf{PIA} & $\bullet$ & & & $\bullet$ & & \\
\textbf{MEA} & $\bullet$ & $\bullet$ & $\bullet$ & $\bullet$ & & $\bullet$ \\ 
\textbf{PA} & $\bullet$ & $\bullet$ & $\bullet$ & $\bullet$ & $\bullet$& \\
\hline
\end{tabular}}
\caption{\textbf{Overview of attack methodologies for each type of attack.} We classify the six attack types across six dimensions of Learning, GAN, Gradient-based, Shadow Training, Threshold, and Equation-solving. Our main objective is to elucidate how learning technology utilizes model parameters and gradients as inputs for the classifier, or employs both input and output queries, to develop alternative models.}
\end{table}

%\begin{figure}[!ht]
%     \centering
%     \includegraphics[width=1\linewidth]{Picture/Fig3.pdf}
%     \caption{Number of libraries for each attack. We use abbreviations to represent the six attack types: Membership Inference Attacks (MIA), Adversarial Attacks (AA), Reconstruction Attacks (RA), Property Inference Attacks (PIA), Model Extraction Attacks (MEA), and Poisoning Attacks (PA). We categorize libraries as original library if they were developed by the authors of the paper, unoriginal library if provided by other individuals or entities, and N/A if the source of the library's origin cannot be determined.}
%     \label{Fig9}
%\end{figure}
\begin{figure}
  \centering
  \begin{tikzpicture}[scale=0.67]
    \begin{axis}[
      every axis plot post/.style={/pgf/number format/fixed},
      ybar,
      bar width=11pt,
      width=\textwidth,
      height=7cm,
       axis on top,
      symbolic x coords={MIA,AA,RA,PIA,MEA,PA},
      xtick=data,
      ylabel=\textbf{{Number of Attack Type}}, % Swap xlabel and ylabel
      xlabel=\textbf{{Attack Type}}, % Swap xlabel and ylabel
      x=2cm,
      ymin=0,
      ymax=15,
    visualization depends on=rawy\as\rawy,
    after end axis/.code={
            \draw [ultra thick, white, decoration={snake, amplitude=1pt}, decorate] (rel axis cs:0,1.0) -- (rel axis cs:1,1.0);
        },
    nodes near coords={
            \pgfmathprintnumber{\rawy}
        },
    every node near coord/.append style={font=\tiny},
    axis lines*=left,
    legend columns=3,
    clip=false,
    legend image code/.code={
      \draw[#1] (0cm,-0.1cm) rectangle (0.4cm,0.1cm);
    },
    legend style={at={(0.05,1)},anchor=north west}
      ]
      \addplot[pattern = horizontal lines, pattern color=red] coordinates {(PA,4) (MEA,6) (PIA,8) (RA,9) (AA,12) (MIA,8)};
      \addplot[pattern = horizontal lines, pattern color=orange] coordinates {(PA,0) (MEA,2) (PIA,0) (RA,3) (AA,3) (MIA,3)};
      \addplot[pattern = horizontal lines, pattern color=blue] coordinates {(PA,2) (MEA,0) (PIA,0) (RA,2) (AA,1) (MIA,2)};
      \legend{Original,Unoriginal,N/A}
    \end{axis}
  \end{tikzpicture}
  \caption{\textbf{Number of libraries for each attack.} Membership Inference Attacks (MIA), Adversarial Attacks (AA), Reconstruction Attacks (RA), Property Inference Attacks (PIA), Model Extraction Attacks (MEA), and Poisoning Attacks (PA). We categorize libraries as original library if they were developed by the authors of the paper, unoriginal library if provided by other individuals or entities, and N/A if the source of the library's origin cannot be determined.}
  \label{fig:vertical_bar_chart}
\end{figure}
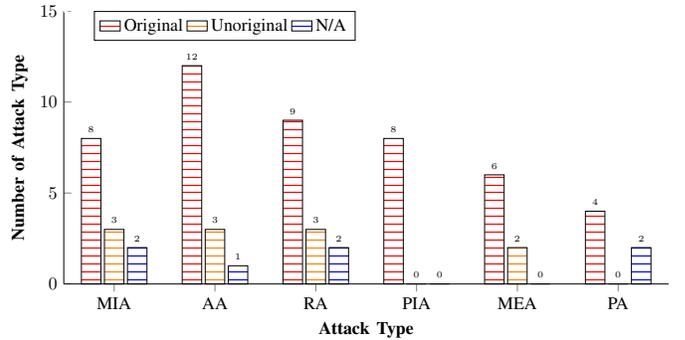

\section{\textbf{Taxonomy of Attacks in ML-based IoT}}
In this section, we delve into the intricacies of six prominent attacks on ML-based IoT systems. We will explore the mechanisms behind each attack and highlight recent advancements that have been able to penetrate these systems successfully. By understanding these vulnerabilities and the current state of the art in exploiting them, we can pave the way for more secure and resilient ML-based IoT applications in the future.
\subsection{\textbf{Membership Inference Attacks}}\label{MIA}
\begin{figure}[!ht]
     \centering
     \includegraphics[width=0.50\textwidth]{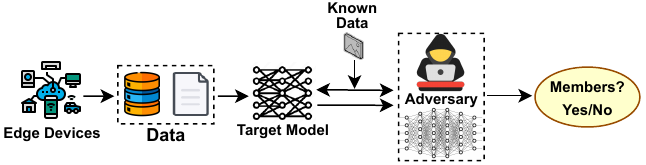}
     \caption{\textbf{Overview of membership inference attacks in ML-based IoT system.}}
     \label{Fig1}
\end{figure}

Membership Inference Attacks have emerged as a noteworthy concern. At the core of membership inference attacks is the attacker's ability to infer whether a specific data point was part of the training dataset used to train an ML model. A visual representation of this process can be found in Fig. \ref{Fig1}. In other words, membership inference attacks are based on methods in which the target model learns features in the training dataset. The attacker constructs his training dataset and observes the model's output on these samples to infer whether the samples in the test dataset belong to the membership samples. This might sound innocuous at first, but consider scenarios where the training data contains sensitive information — revealing whether a particular data point (like a patient's medical record) was used in training could lead to significant privacy breaches. 

One of the seminal works in this domain was presented by Shokri et al. \cite{shokri2017membership}. They provided a concrete framework to understand membership inference attacks, introducing an attack model \( f_{\text{attack}} \). The input $x_{attack}$ is a prediction confidence vector consisting of the correct label class and a target model (the model being attacked). The output of the attack model is a prediction class “in” (member) or “out” (non-member).
\begin{flalign}
& f_{attack}(x_{attack}) \rightarrow y\\    
& x_{attack}=(label,prediction),y \in \{in,out\} 
\end{flalign}

%The membership inference attack process can be divided into two distinct phases. In the first phase, an attacker selects the target model and predicts it on a test dataset, ensuring it's distinct from the model's training dataset. During the second phase, the attacker opts for an appropriate membership inference algorithm. From a broad dataset, a subset is chosen to train the attack model. Subsequently, the attacker earmarks a small fraction of the target model's training dataset as membership samples. The purpose of these samples is to train an attack model that can differentiate between member and non-member samples. Using this trained attack model, the attacker can then ascertain if a prediction result aligns with a member sample. 
For further insights, we show the experimental results in TABLE \ref{tablemia} and briefly introduce seminal and contemporary works on membership inference attacks. Key references include \cite{shokri2017membership,long2018understanding,yeom2018privacy,rahman2018membership,song2019privacy,sablayrolles2019white,wang2019beyond,yaghini2019disparate,truex2019demystifying,irolla2019demystifying,song2019membership,truex2019effects,long2020pragmatic,leino2020stolen,humphries2020differentially,shokri2021privacy,chen2021machine,hui2021practical,song2021systematic,hu2021source,choquette2021label,liu2022membership,carlini2022membership,ye2022enhanced,van2023membership,yuan2023interaction,matsumoto2023membership,bertran2023scalable,zhang2023mida,hu2023membership}.  (For some unintroduced papers, we also display them in TABLE \ref{tab}).

Shokri et al. \cite{shokri2017membership} propose a method to determine if a record belongs to the model's training dataset. They achieve this by providing black-box access to the target model and using shadow model training techniques to mimic the target model's behavior. To counter such attacks, they suggest mitigation strategies, including restricting the prediction vector to the top $k$ classes, coarsening the precision of the prediction vector, increasing the entropy of the prediction vector, and employing regularization.
Rong et al. \cite{salem2018ml} take a different approach by training a single shadow model with various datasets to capture the membership state of data points in the ML training set. They propose combating broader membership inference attacks by employing techniques such as dropout and model stacking for defense.
Long et al. \cite{long2018understanding} identify individual records with high accuracy in real-world datasets using black-box queries on ML models. They argue that combining generalization and perturbation with careful training set selection can be used to defend against such attacks.
Nasr et al. \cite{nasr2018comprehensive} assess the privacy vulnerability of the stochastic gradient descent algorithm and design a deep learning attack model to calculate the membership probability of target data points. They also evaluate the privacy of pre-trained state-of-the-art models on the CIFAR-100 dataset to validate the effectiveness of white-box inference attacks.
Liwei Song et al. \cite{song2019privacy} propose two new membership inference attacks against adversarial robust models, leveraging predictions from adversarial examples and validated worst-case predictions. Their core approach exploits the structural characteristics of adversarial robust defense to achieve high inference accuracy. They further validate the effectiveness of these attacks against six state-of-the-art adversarial defense methods.
Choquette et al. \cite{choquette2021label} introduce a labeling attack that assesses the robustness of a model's predicted labels to input perturbations to infer membership. They also confirm that training with differential privacy \cite{abadi2016deep} or strong L2 regularization \cite{nasr2018machine} can serve as an effective defense.
Song et al. \cite{song2021systematic} measure the privacy risk of a target model using improved attacks with class-specific threshold settings and a new inference attack based on enhanced prediction entropy estimates. They demonstrate that these methods are not as effective as previous attacks when tested against defense mechanisms like adversarial regularization \cite{nasr2018machine} and MemGuard \cite{jia2019memguard}.
BLINDMI \cite{hui2021practical} introduces a practical approach where existing samples are transformed into new samples to create a dataset with non-membership. Samples are differentially moved from the target dataset to the generated non-membership set, and membership status is determined based on the distance moved by the sample. Hui et al. verify that BLINDMI can break these defenses by achieving reasonable F1-score under different privacy utility budgets against adversarial regularization \cite{nasr2018machine}, MemGuard \cite{jia2019memguard}, Mixup+MMD \cite{li2020membership}, and differential privacy \cite{abadi2016deep}.
A Bayesian perspective is adopted by Hu et al. \cite{hu2021source} to demonstrate that honest but curious servers can initiate Source Information Attacks (SIA) to steal non-trivial source information of training members without violating the federated learning protocol. The server exploits the prediction loss of the local model on the training members to implement the attack and evaluates the key factors in SIA on six datasets to verify its effectiveness.
Knowledge extraction \cite{hinton2015distilling} is employed as an attack method by Liu et al. \cite{liu2022membership}. The adversary evaluates the loss trajectory of the target sample on each intermediate extraction model to obtain the loss trajectory of the target sample. Finally, the attack model takes the extraction loss trajectory of the target sample and its loss on the original target model as input to infer the membership status of the sample. They further validate that the attack exhibits advantages in different indicators across various datasets and model architectures.

\begin{table*}[!ht]
\caption{\textbf{Reviews of membership inference attack libraries.} The library details are presented in the table below. In this table, ``Ref." refers to the cited paper, or is marked N/A if no corresponding paper was located in our library. ``Attack Type" describes the kind of attack in the paper, and ``White/Black-box" identifies the attack setting. ``Library Link" offers a link to the mentioned library, while ``Original" specifies if the library was created by the authors of the paper. ``Unoriginal Library" pertains to libraries sourced from other individuals or entities, or N/A if the origin remains unclear. ``Impl." indicates whether the main program of the library was run successfully after debugging. The version of Python utilized during debugging is mentioned, as well as if a GPU was used. ``Dataset" names the dataset the library relies on, while ``Evaluation" displays the debugging results or is marked N/A if results cannot be shown due to being in image format or if debugging was unsuccessful. ``Issues" pinpoint primary challenges faced during debugging, such as missing essential library files termed ``Lost files." ``Debugging solution" lists the adjustments made during debugging, ``Debugging failed" notes when severe bugs obstructed the debugging process, and ``Environment loading failed" is used when setting up a working runtime environment was unattainable, or N/A if debugging proceeded without significant issues. It's vital to recognize that this description pertains to tables from TABLE. \ref{table1} through TABLE. \ref{table5}. However, TABLE. \ref{table6} (focused on poisoning attacks) omits the ``White/Black-box" column. For clarity, in its first entry, we employed the original library [L.1] for data attacks in a black-box setting, utilizing Python 3.10 and the MNIST dataset. Post-debugging, the outcomes are documented in the evaluation.} 
\label{tablemia}
\begin{center}\scriptsize
\hspace{-30pt}
\begin{tabular}{ccccccccccc}\hline
% Ref&Attack Type&White/Black-Box&\makecell[c]{Library Link\\
% (Appendix)}&\makecell[c]{Original Library}&Implementation&Python&GPU&Dataset&Evaluation&Issues\\ 
% \hline
\textbf{Ref.} & \textbf{Attack Type} & \textbf{White/Black-Box} & \textbf{\makecell[c]{Library Link \\ (Appendix)}} & \textbf{\makecell[c]{Original}} & \textbf{Impl.} & \textbf{Python} & \textbf{GPU} & \textbf{Dataset} & \textbf{Evaluation} & \textbf{Issues} \\
\hline\hline

\multirow{8}{*}{\makecell[c]{\cite{shokri2017membership}}}&\multirow{8}{*}{\makecell[c]{Data Attack}} &\multirow{8}{*}{\makecell[c]{Black-Box}}&\multirow{1}{*}{[L.1]} &\multirow{1}{*}{True} &\multirow{1}{*}{True} &\multirow{1}{*}{{3.10}} &\multirow{1}{*}{False} &\multirow{1}{*}{MNIST}&\makecell[c]{ Training Accuracy: 0.59\\Testing Accuracy: 0.57}&\makecell[c]{Lost files/\\Debugging \\solution} \\ 
\cline{4-11}

{}&{}&{}&{\makecell[c]{[L.2]}}&{\makecell[c]{False}}&{\makecell[c]{True}}&{\makecell[c]{3.7}}&{\makecell[l]{False}}&CIFAR-10&{\makecell[c]{Precision: 0.72\\
Recall: 0.71\\
F1-Score: 0.71\\
Accuracy: 0.71
}}
&{\makecell{Debugging \\solution}}\\
\cline{4-11}

{}&{}&{}&{\makecell[c]{[L.3]}}& {\makecell[c]{False}}&{\makecell[c]{True}}&{\makecell[c]{3.10}}&{\makecell[l]{False}}&CIFAR-10&{\makecell[c]{Precision: 0.54\\
Recall: 0.90\\
F1-Score: 0.67\\
Accuracy: 0.56}}
&{\makecell{N/A}}\\
\hline

\multirow{3}{*}{\makecell[c]{\cite{salem2018ml}}}  &\multirow{3}{*}{\makecell[c]{Data Attack}} &\multirow{3}{*}{\makecell[c]{Black-Box}}&\multirow{1}{*}{[L.4]} &\multirow{1}{*}{True} &\multirow{1}{*}{False} &\multirow{1}{*}{{2.7}} &\multirow{1}{*}{False} &\multirow{1}{*}{CIFAR-10}&\makecell[c]{N/A}&\makecell[c]{Lost files} \\ 
\cline{4-11}
 
{}&{}&{}&{\makecell[c]{[L.5]}}&{\makecell[c]{False}}&{\makecell[c]{True}}&{\makecell[c]{3.10}}&{\makecell[l]{False}}&CIFAR-10&{\makecell[c]{Accuracy: 59.43\\Training Loss: 135.30}}&{N/A}\\
\hline

\multirow{7}{*}{\makecell[c]{\cite{song2019privacy}}}&\multirow{7}{*}{\makecell[c]{Data Attack}} &\multirow{7}{*}{\makecell[c]{Black-Box}}&\multirow{7}{*}{[L.6]} &\multirow{7}{*}{True} &\multirow{7}{*}{True} &\multirow{7}{*}{3.7} &\multirow{7}{*}{False} &\multirow{1}{*}{Yale}&\makecell[c]{ Training Accuracy: 0.94\\
Testing Accuracy: 0.69\\
Attack Accuracy: 0.66}&{\makecell{Debugging \\solution}}\\
\cline{9-11}

{}&{}&{}&{}&{}&{}&{}&{}&{\makecell{Fashion-\\MNIST}}&{\makecell[c]{Training Accuracy: 0.83\\
Testing Accuracy: 0.78\\
Attack Accuracy: 0.52
}}&{\makecell{Debugging \\solution}}\\
\cline{9-11}

{}&{}&{}&{}&{}&{}&{}&{}&{\makecell{CIFAR-10}}&{\makecell[c]{Training Accuracy: 0.76\\
Testing Accuracy: 0.47\\
Attack Accuracy: 0.67
}}&{\makecell{Debugging \\solution}}\\
\hline

\cite{song2021systematic}&\multicolumn{1}{c}{\multirow{1}*{Model Attack}}&{\makecell[c]{Black-Box}}  &{\makecell[c]{[L.7]}}&{\makecell[c]{True}}&{\makecell[c]{True}}&{\makecell[c]{3.10}}&{\makecell[c]{False}}&Location&{\makecell[c]{Training Accuracy: 1.00\\
Testing Accuracy: 0.63\\
Attack Accuracy: 0.69}}&{\makecell{N/A}}\\
\hline

\cite{jia2019memguard}&\multicolumn{1}{c}{\multirow{1}*{Model Attack}}&{\makecell[c]{Black-Box}}  &{\makecell[c]{[L.8]}}&{\makecell[c]{True}}&{\makecell[c]{True}}&{\makecell[c]{2.7}}&{\makecell[c]{False}}&Location&{\makecell[c]{Attack Accuracy: 0.50\\
}}&{\makecell{Debugging \\solution}}\\
\hline

\cite{hui2021practical}&\multicolumn{1}{c}{\multirow{1}*{Data Attack}}&{\makecell[c]{Black-Box}}   &{\makecell[c]{[L.9]}}&{\makecell[c]{True}}&{\makecell[c]{True}}&{\makecell[c]{3.10}}&{\makecell[c]{True}}&CIFAR-10&{\makecell[c]{Precision: 0.83\\ 
Recall: 0.63\\
F1-Score: 0.72\\
Accuracy: 0.75\\
}}&{\makecell{N/A}}\\
\hline

\multirow{3}{*}{\makecell[c]{\cite{hu2021source}}}  &\multirow{3}{*}{\makecell[c]{Data Attack}} &\multirow{3}{*}{\makecell[c]{Black-Box}}&\multirow{3}{*}{[L.10]} &\multirow{3}{*}{True} &\multirow{3}{*}{True} &\multirow{3}{*}{{3.10}} &\multirow{3}{*}{True} &\multirow{1}{*}{Syntheic}&\makecell[c]{Training Accuracy: 0.93\\
Testing Accuracy: 0.92\\
}&{\makecell{N/A}}\\ 
\cline{9-11}
 
{}&{}&{}&{}&{}&{}&{}&{}&MNIST&{\makecell[c]{Training Accuracy: 1.00\\
Testing Accuracy: 0.99\\
}}&{\makecell{Debugging \\solution}}\\
\hline

\cite{liu2022membership}&\makecell[l]{Model Attack}&{\makecell[c]{Black-Box}}   &{\makecell[c]{[L.11]}}&{\makecell[c]{True}}&{\makecell[c]{True}}&{\makecell[c]{3.10}}&{\makecell[l]{True}}&{\makecell[l]{CIFAR-100}}&{\makecell[c]{Recall: 0.10\\
FPR: 0.40\\
Max AUC: 0.59\\
Attack Accuracy: 0.55\\
}}&{\makecell{Debugging \\solution}}\\
\hline

\multirow{3}{*}{\makecell[c]{N/A}}  &\multirow{3}{*}{\makecell[c]{Data Attack}} &\multirow{3}{*}{\makecell[c]{Black-Box}}&\multirow{3}{*}{[L.12]} &\multirow{3}{*}{N/A} &\multirow{3}{*}{True} &\multirow{3}{*}{{3.7}} &\multirow{3}{*}{True} &\multirow{1}{*}{MNIST}&\makecell[c]{Attack Accuracy: 0.53\\
Testing Accuracy: 0.52\\
}&{\makecell{Debugging \\solution}}\\
\cline{9-11}
 
{}&{}&{}&{}&{}&{}&{}&{}&CIFAR-10&{\makecell[c]{Attack Accuracy: 0.74\\
Testing Accuracy: 0.73\\
}}&{\makecell{Debugging \\solution}}\\
\hline

{N/A}&\makecell[l]{Model Attack}&{\makecell[c]{Black-Box}}&{\makecell[c]{[L.13]}}&{\makecell[c]{N/A}}&{\makecell[c]{True}}&{\makecell[c]{3.7}}&{\makecell[c]{False}}&CIFAR-10&{\makecell[c]{Training Accuracy: 0.76\\
Testing Accuracy: 0.63\\
}}&{\makecell{Debugging \\solution}}\\
\hline

\end{tabular}
\label{table1} 
\end{center}
\end{table*}

\subsection{\textbf{Adversarial Attacks}}\label{AA}
\begin{figure}[!ht]
     \centering
     \includegraphics[width=0.50\textwidth]{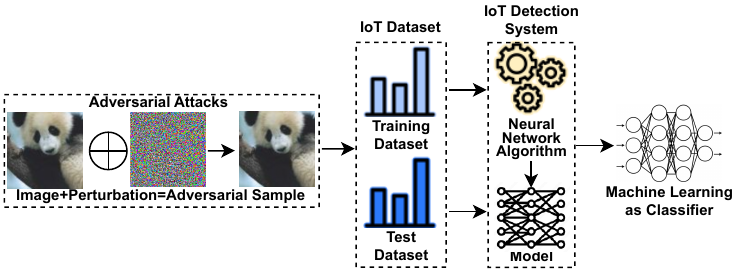}
     \caption{\textbf{Overview of adversarial attacks in ML-based IoT system.}}
     \label{Fig2}
\end{figure}

Adversarial attack is based on the model's vulnerability and the sample's small perturbation. Attackers cheat or mislead the output of a model by making small, targeted perturbations to the input data. The process of adversarial attacks is as shown in Fig. \ref{Fig2}.

Szegedy et al. \cite{szegedy2013intriguing} by finding the smallest addition to the loss function, the neural network makes misclassification and transforms the problem into a convex optimization. where $f: \mathbb{R}^m \rightarrow \lbrace 1 . . .k \rbrace$ represents a classifier that maps image pixel value vectors to discrete label sets. In addition, it is also assumed that $f$ has an associated continuous loss function, denoted as $loss_f: \mathbb{R}^m \times \lbrace 1 . . .k \rbrace \rightarrow \mathbb{R}^+$. For a given $x \in \mathbb{R}^m$ image and target label $l\in \lbrace 1 . . .k \rbrace.$ The goal is to solve the following box-constrained optimization problem:\\
\centerline{Minimize $||r||_2$ subject to:}
\vspace{-0.5cm}
\begin{flalign}
& f(x+r)=l \\
& x+r \in [0,1]^m
\end{flalign}
Since the minimum $r$ may not be unique, For an arbitrarily chosen minimizer $D(x,l)$, denoted by $x + r.$ $D(x,L)$ can be found by row search to find the minimum value of $c > 0$, so that the minimum value of r satisfies $f(x+r) = l.$
\begin{flalign}
\text{Minimize } c|r| + loss_f (x + r, l) \notag\\
\text{ subject to }  x + r \in [0,1]^m
\end{flalign}

%Generally, the process is divided into two phases. The first phase is when the attacker selects the target model and makes predictions on the generated adversarial samples.
%The second phase is that the attacker chooses a suitable adversarial attack algorithm. An input sample is then selected from the dataset as the target of the attack. Generate adversarial samples using the selected attack algorithm and perturb the samples to deceive the target model. Input the generated adversarial samples into the target model for prediction and obtain the prediction results of the target model for the adversarial samples. 

In this section, we show the experimental results in TABLE \ref{tableaa} and briefly introduce some early classic articles and current cutting-edge work on adversarial attacks selected from \cite{szegedy2013intriguing,goodfellow2014explaining,moosavi2016deepfool,kurakin2016adversarial,madry2017towards,huang2017adversarial,carlini2017towards,chen2017zoo,papernot2017practical,brendel2017decision,zugner2018adversarial,dong2018boosting,ilyas2018black,bojchevski2019adversarial,guo2019simple,su2019one,chen2020hopskipjumpattack,zugner2020adversarial,bin2020adversarial,wang2021admix,rony2021augmented,amich2022eg,jia2022adversarial,li2022subspace,xiong2022stochastic,song2023robust,kim2023demystifying,feng2023dynamic,wang2023towards,chen2023effective,huang2023t}.  (For some unintroduced papers, we also display them in TABLE \ref{tab}).

Goodfellow et al. \cite{goodfellow2014explaining} introduced a straightforward method for generating adversarial samples known as FGSM (Fast Gradient Sign Method), and further utilized these adversarial samples for adversarial training, demonstrating that such training can lead to regularization and offer more significant regularization benefits compared to dropout.
Moosavi et al. \cite{moosavi2016deepfool} proposed the DeepFool algorithm, designed to calculate perturbations capable of deceiving deep neural networks. DeepFool leverages iterative linearization within a classifier-based framework to converge upon the smallest perturbations necessary to misclassify an input image, thereby finding the minimal distortion required for misclassification.
Kurakin et al. \cite{kurakin2016adversarial} extended adversarial training to the ImageNet dataset, building upon the earlier FGSM attack method \cite{goodfellow2014explaining}. This extension, referred to as "basic iterative methods," primarily focuses on reducing the loss function of the target category during training.
Carlini et al. \cite{carlini2017towards} leveraged the internal configuration of the target deep neural network (DNN) for attack bootstrapping. They quantified the difference between the adversarial attack and the original example using the $L_2$ norm. Additionally, they validated that their method could successfully defeat defensive distillation \cite{hinton2015distilling}.
Nicolas Papernot et al. \cite{papernot2017practical} devised a technique that involves training a local model to replace the target DNN. This local model is trained using inputs synthesized by the adversary and labeled by the target DNN. Adversarial examples are crafted by making local substitutions to identify inputs misclassified by the target DNN. Furthermore, they explore two defense strategies: Adversarial training \cite{goodfellow2014explaining,szegedy2013intriguing,wang2017two} and Defensive distillation \cite{hinton2015distilling}.
Chen et al. \cite{chen2017zoo} introduced an attack method that accesses both the input and output of the target DNN. This zero-order optimization-based attack directly estimates the gradient of the target DNN to generate adversarial examples, and uses techniques such as zero-order random coordinate descent, dimensionality reduction, hierarchical attacks, and importance sampling to effectively attack black-box models. They also discuss various defense \cite{feinman2017detecting,grosse2017statistical,huang2017safety,metzen2017detecting,xu2017feature} strategies, including detection-based defenses, gradient and representation masking, as well as adversarial training.\cite{madry2017towards,jin2015robust,tramer2017ensemble,zantedeschi2017efficient,zheng2016improving}.
Ilyas et al. \cite{ilyas2018black} defined three threat models: query limit setting, partial information setting, and label-only setting. They proposed the use of NES (Nesterov's Estimation of the Stochastic Gradient)\cite{nesterov2017random} as a black-box gradient estimation technique, along with PGD (Projected Gradient Descent) using estimated gradients for white-box attacks to construct adversarial examples. The effectiveness of this approach was demonstrated on ImageNet classifiers under these three threat models.
Su et al. \cite{su2019one} generates a single-pixel adversarial perturbation based on differential evolution, only modifies one pixel to attack the image, and deceives more types of networks. They show that non-target attacks can be launched with considerable accuracy by modifying only one pixel on three common deep neural network structures.
Chen et al. \cite{chen2020hopskipjumpattack} propose the HopSkipJumpAttack algorithm based on the decision boundary using bipartite information to estimate the gradient direction, which can approximate the gradient in the black box case, thereby generating targeted and non-targeted adversarial samples. In addition, they verified the defense methods of defensive distillation \cite{hinton2015distilling}, region-based classification \cite{cao2017mitigating}, and adversarial training \cite{goodfellow2014explaining,kurakin2016adversarial,madry2017towards,tramer2017ensemble} to show that the attack is competitive.
Wang et al. \cite{wang2022triangle} propose a novel method Triangle Attack (TA) based on the law of sines, which exploits the geometric information related to the larger angles in any triangle by optimizing perturbations. With TA, effective dimensionality reduction can be achieved in the low-frequency space, and compared to recent work \cite{cheng2018query,cheng2019sign,chen2020hopskipjumpattack,li2020qeba,rahmati2020geoda,maho2021surfree}, the same attack success rate can be achieved with fewer queries.

%\clearpage
\begin{table*}[!ht]
\caption{\textbf{Reviews of adversarial attack libraries.} Taking the first line as an example, we utilized the original library [L.14] to conduct data attacks in a black-box environment. We used Python 3.10, with GPU, and the ImageNet dataset. After debugging, the results are shown in the evaluation. Additionally, it's worth noting that the final evaluation of this library is based on the output images, and we only present the program's running results.}
\label{tableaa}
\begin{center}\scriptsize
\begin{tabular}{ccccccccccc}\hline
% Ref.&Attack Type&White/Black-Box&\makecell[c]{Library Link\\
% (Appendix)}&\makecell[c]{Original Library}&Implementation&Python&GPU&Dataset&Evaluation&Issues\\ 
% \hline
\textbf{Ref.} & \textbf{Attack Type} & \textbf{White/Black-Box} & \textbf{\makecell[c]{Library Link \\ (Appendix)}} & \textbf{\makecell[c]{Original}} & \textbf{Impl.} & \textbf{Python} & \textbf{GPU} & \textbf{Dataset} & \textbf{Evaluation} & \textbf{Issues} \\
\hline\hline
\cite{moosavi2016deepfool}&Data Attack&{\makecell[c]{Black-Box}}&[L.14]&True&True&3.10&True&\makecell[c]{ImageNet}&\makecell[c]{Original label:\\n03538406 horse cart\\
Perturbed label:\\n02437312 Arabian camel}&{\makecell{Debugging \\solution}}\\
\hline

\multirow{3}{*}{\cite{carlini2017towards}} &\multirow{3}{*}{\makecell[c]{Model Attack}} &\multirow{3}{*}{\makecell[c]{Black-Box}}&\multirow{3}{*}{\makecell[c]{[L.15]}} &\multirow{3}{*}{\makecell[c]{True}} &\multirow{3}{*}{\makecell[c]{True}} &\multirow{3}{*}{\makecell[c]{3.10}} &\multirow{3}{*}{\makecell[c]{True}} &{CIFAR-10} &\makecell[c]{L0 Total Distortion: 2.34\\
L2 Total Distortion: 0.21}&\makecell[c]{Debugging \\solution} \\ 
\cline{9-11}
 {}&{}&{}&{}&{}&{}&{}&{}&{MNIST}&{\makecell[c]{L0 Total Distortion: 5.00\\
L2 Total Distortion: 2.02}}
&{\makecell{Debugging \\solution}}\\
\hline

\cite{chen2017zoo}&Model Attack&{\makecell[c]{Black-Box}}&{\makecell[c]{[L.16]}}& {\makecell[c]{True}}&{\makecell[c]{True}}&{\makecell[c]{3.7}}&{\makecell[c]{True}}&{\makecell[c]{MNIST}}&{\makecell[c]{Success = True\\
Const = 0.325\\
Distortion = 1.80\\
Success\_rate = 1.00\\
L2\_avg = 2.34}}
&{\makecell{Debugging \\solution}}\\
\hline

\cite{guo2019simple}&\makecell[c]{Model Attack}&{\makecell[c]{Black-Box}}&{\makecell[c]{[L.17]}}& {\makecell[c]{True}}&{\makecell[c]{True}}&{\makecell[c]{3.10}}&{\makecell[c]{True}}&{\makecell[c]{MNIST}}&{\makecell[c]{Accuracy: 0.99}}
&{\makecell{Debugging \\solution}}\\
\hline

\multirow{3}{*}{\makecell[c]{\cite{du2019query}}}&\multirow{3}{*}{\makecell[c]{Model Attack}} &\multirow{3}{*}{\makecell[c]{Black-Box}}&\multirow{3}{*}{\makecell[c]{[L.18]}} &\multirow{3}{*}{\makecell[c]{True}} &\multirow{3}{*}{\makecell[c]{True}} &\multirow{3}{*}{\makecell[c]{3.10}} &\multirow{3}{*}{\makecell[c]{True}} &{CIFAR-10} &\makecell[c]{Avg\_query = 963.38\\
Accuracy = 1.00}&\makecell[c]{Debugging \\solution} \\ 
\cline{9-11}
{}&{}&{}&{}&{}&{}&{}&{}&{\makecell[c]{CIFAR-10}}&{\makecell[c]{Avg\_query = 1726.11\\
Accuracy = 0.48}}
&{\makecell{Debugging \\solution}}\\
\hline

\cite{wang2022triangle}&{Model Attack}&{\makecell[c]{Black-Box}}&{\makecell[c]{[L.19]}}& {\makecell[c]{True}}&{\makecell[c]{True}}&{\makecell[c]{3.10}}&{\makecell[c]{True}}&{\makecell[c]{Image(200)}}&{\makecell[c]{Top-1 Accuracy: 1.00\\
Top-2 Accuracy: 0.93\\
Top-3 Accuracy: 0.51}}
&{\makecell{N/A}}\\
\hline

\cite{goodfellow2014explaining}&\multirow{3}{*}{Model Attack}&\multirow{3}{*}{Black-Box}&\multirow{3}{*}{[L.20]}&\multirow{3}{*}{False}&\multirow{3}{*}{True}&\multirow{3}{*}{3.10}&\multirow{3}{*}{True}&\multirow{3}{*}{\makecell[c]{Fashion-\\MNIST}}&{\makecell[c]{Accuracy: 0.95}}
&{\makecell{Debugging \\solution}}\\
\cline{1-1}
\cline{10-11}

\cite{wang2021adversarial}&{}&{}&{}&{}&{}&{}&{}&{}&{\makecell[c]{Accuracy : 0.93}}
&{\makecell{Debugging \\solution}}\\
\hline
 
\cite{rauber2020foolbox}&\makecell[c]{Model Attack}&{\makecell[c]{Black-Box}}&{\makecell[c]{[L.21]}}& {\makecell[c]{True}}&{\makecell[c]{True}}&{\makecell[c]{3.10}}&{\makecell[c]{True}}&{\makecell[c]{ImageNet}}&{\makecell[c]{Clean Accuracy:  0.94\\
Attack Success:0.63\\ 
Robust Accuracy:0.38}}
&{\makecell{N/A}}\\
\hline

\cite{amich2022eg}&\makecell[c]{Model Attack}&{\makecell[c]{Black-box/\\White-Box}}&{\makecell[c]{[L.22]}}& {\makecell[c]{True}}&{\makecell[c]{True}}&{\makecell[c]{3.10}}&{\makecell[c]{True}}&{\makecell[c]{CIFAR-10}}&{\makecell[c]{Current EG-Booster \\Evasion rate on FGS: 0.15\\
Evasion rate of\\ EG-Booster attack: 0.15\\
Max change :-0.4925\\
Avg change :-0.4975\\
Avg unperturbed\\ positives :15.1719}}
&{\makecell{Debugging \\solution}}\\
\hline

\cite{brendel2017decision}&\makecell[c]{Model Attack}&{\makecell[c]{Black-Box}}&{\makecell[c]{[L.23]}}& {\makecell[c]{False}}&{\makecell[c]{True}}&{\makecell[c]{3.7}}&{\makecell[c]{False}}&{\makecell[c]{ImageNet}}&{\makecell[c]{Mean Squared\\Error: 756.6158}}
&{\makecell{Debugging \\solution}}\\
\hline

\cite{miyato2015distributional}&\makecell[c]{Model Attack}&{\makecell[c]{Black-Box}}&{\makecell[c]{[L.24]}}& {\makecell[c]{False}}&{\makecell[c]{True}}&{\makecell[c]{3.7}}&{\makecell[c]{False}}&{\makecell[c]{MNIST}}&{\makecell[c]{Test Error Rate:0.369}}
&{\makecell{Debugging \\solution}}\\
\hline

\cite{su2019one}&\makecell[c]{Model Attack}&{\makecell[c]{Black-Box}}&{\makecell[c]{[L.25]}}& {\makecell[c]{True}}&{\makecell[c]{True}}&{\makecell[c]{3.10}}&{\makecell[c]{True}}&{\makecell[c]{CIFAR-10}}&{\makecell[c]{Confidence: 0.075\\
Prior confidence 0.501\\
Attack success: True}}
&{\makecell{Debugging \\solution}}\\
\hline
 
\cite{pajola2021fall}&\makecell[c]{Data Attack}&{\makecell[c]{Black-Box}}&{\makecell[c]{[L.26]}}& {\makecell[c]{True}}&{\makecell[c]{True}}&{\makecell[c]{3.10}}&{\makecell[c]{True}}&{\makecell[c]{Hatefui\\Memes}}&{\makecell[c]{Negative Scores:\\
Original: 0.61\\
Mask1: 0.00\\
Mask2: 0.00}}
&{\makecell{N/A}}\\
\hline

\cite{ilyas2018black}&\makecell[c]{Model Attack/\\Data Attack}&{\makecell[c]{Black-Box}}&{\makecell[c]{[L.27]}}& {\makecell[c]{True}}&{\makecell[c]{True}}&{\makecell[c]{3.10}}&{\makecell[c]{True}}&{\makecell[c]{ImageNet}}&{\makecell[c]{Loss: 1.7815\\Lr : 3.13E-04\\ Eps: 0.050}}
&{\makecell{Debugging \\solution}}\\
\hline

\cite{chen2020hopskipjumpattack}&\makecell[c]{Model Attack}&{\makecell[c]{Black-Box}}&{\makecell[c]{[L.28]}}& {\makecell[c]{True}}&{\makecell[c]{True}}&{\makecell[c]{3.10}}&{\makecell[c]{True}}&{\makecell[c]{CIFAR-10}}&{\makecell[c]{Accuracy: 0.9244\\L2 distance: 1.0344E+00}}
&{\makecell{Debugging \\solution}}\\
 \hline
 
{N/A}&Model Attack&{\makecell[c]{Black-Box}}&{\makecell[c]{[L.29]}}& {\makecell[c]{N/A}}&{\makecell[c]{True}}&{\makecell[c]{3.7}}&{\makecell[c]{False}}&{\makecell[c]{MNIST}}&{\makecell[c]{Testing Accuracy: 0.99}}
&{\makecell{Debugging \\solution}}\\
\hline
 
\end{tabular}
\label{table2} 
\end{center}
\end{table*}

\subsection{\textbf{Reconstruction Attacks}}\label{RA}
\begin{figure}[!ht]
     \centering
     \includegraphics[width=0.50\textwidth]{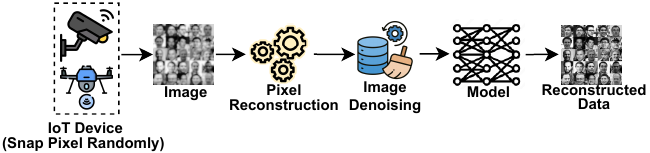}
     \caption{\textbf{Overview of reconstruction attacks in ML-based IoT system.}}
     \label{Fig3}
\end{figure}

Reconstruction attacks are based on model output data. The attacker reconstructs or restores some features of the input data by analyzing the model output and known information. The process of reconstruction attacks is as shown in Fig. \ref{Fig3}.

The environment for the reverse attack of the model proposed by Fredrikson el al. \cite{fredrikson2015model} is mainly divided into black-and-white box conditions.

Start by pinning the tree, as shown below, where $(x,y)$ represents a target instance, which may or may not be in the training set.
\begin{flalign}
f(x)=\sum\nolimits_{i=1}^m \omega_i \phi_i(x)
\end{flalign}

Under black box conditions, the MI Attack algorithm proposed by Fredrikson et al. \cite{fredrikson2015model} was used and fine-tuned for the decision tree scenario. The specific differences are reflected in the following: the target model can only produce discrete outputs; the error model information is different, and the confusion matrix  $C$ is used instead of Gaussian. The variance of the distribution. The definition is as follows:
\begin{flalign}
err(y,y^{'}) \propto Pr \text{ } [f(x)=y^{'} | \text{ } \text{y is the true label]}
\end{flalign}

In the white-box condition, the attacker knows each $\phi_i$, the total number of samples in the training set. 
\begin{flalign}
N=\sum\nolimits_{i=1}^m n_i 
\end{flalign}

The number of training samples $n_i$ (corresponding to $\phi_i$) introduces a series of paths to the known feature vector $x_k:$
\begin{align}
S &= \lbrace {s_i} \rbrace_{1\leq i\leq m}: \notag\\
&\quad S = \lbrace (\phi_i,n_i) |\hspace{0.1cm}\exists x' \in \mathbb{R}^d .\hspace{0.1cm} x'_k=x_k \land \phi_i(x')\rbrace
\end{align}

Each path corresponds to a basis function $\phi_i$ and a sample count $ni.$ Let $pi=ni/N,$ related to the joint distribution used to construct the training set features. Use the basis function $\phi_i$ to divide the feature space; v represents the specific value of the first feature, and $v_k$ represents the specific value of the $d-1$ dimensional non-private feature. Finally, as shown below, a white-box predictor with counting (WBWC) is constructed by outputting $v$ to maximize the formula and return MAP predictions given additional auxiliary information.
\begin{align}
&Pr[{x_1}=v | (s_1 \vee \cdot \cdot \cdot \vee s_m)\wedge x_k =v_k]\notag\\
&\propto \sum_{i=1}^m \frac{p_i \phi_i(v) \cdot Pr [x_k = v_k \cdot Pr [x_1 = v]}{\sum_{j=1}^m p_j \phi_j(v)}\notag\\
&\propto \frac{1}{\sum_{j=1}^m p_j \phi_j(v)}\sum_{1\leq i\leq m} p_i \phi_i(v) \cdot Pr [x_1 = v]
\end{align}

%Generally, the process is divided into two phases. The first phase is that the attacker first inputs data and collects the output data of the target model.
%The second phase is that the attacker uses the reconstruction algorithm to reconstruct or restore the input data's information by analyzing the model's output and known information. The attacker can obtain information about the original data through the reconstructed input data, such as images, texts, etc. 

In this section, we show the experimental results in TABLE \ref{tablera} and briefly introduce some early classic articles and current cutting-edge work on reconstruction attacks selected from \cite{fredrikson2014privacy,fredrikson2015model,wu2016methodology,hitaj2017deep,song2017machine,hidano2017model,sannai2018reconstruction,lacharite2018improved,wang2019beyond,carlini2019secret,he2019model,yang2019neural,zhu2019deep,aivodji2019gamin,oh2019exploring,salem2020updates,zhang2020secret,xu2020midas,goldsteen2020reducing,zhu2020r,wang2021variational,zhao2021exploiting,dong2021privacy,fereidooni2021safelearn,struppek2022plug,balle2022reconstructing,kahla2022label,han2023reinforcement,annamalai2023linear,long2023membership}.  (For some unintroduced papers, we also display them in TABLE \ref{tab}).

Fredrikson et al. \cite{fredrikson2015model} utilize the confidence values exposed by an API to infer the pixel values in an image. They achieve this by iteratively testing different pixel intensity values to reconstruct the image. Additionally, they propose countermeasures against such attacks, including the use of decision trees and reducing the quality or accuracy of gradient information retrieved from the model.
In addition, they highlight that, given an ML model and some features of a training sample, an attacker can potentially recover the remaining features of the sample \cite{fredrikson2014privacy}. However, they suggest that carefully selecting a privacy budget can effectively mitigate model inversion attacks using differential privacy mechanisms.
The attack developed by Hitaj et al. \cite{hitaj2017deep} capitalizes on the real-time nature of the learning process and relies on Generative Adversarial Networks (GANs). This attack involves merging generated samples with the attacker's labels into the training set to recover the victim's labeled samples. They suggest that secure multi-party computation or fully homomorphic encryption could potentially counteract this attack.
Zhu et al. \cite{zhu2019deep} use DLG to obtain training inputs and labels simultaneously in several iterations. Full display of private training data (including inputs and labels) after optimization. They also proposes some defense strategies against this attack, such as noisy gradients, gradient compression \cite{lin2017deep,tsuzuku2018variance} and sparsification large batch, high resolution and cryptology \cite{bonawitz2016practical,aono2017privacy}. Without changing the training setting, the most effective defense is gradient pruning.
Wang et al. \cite{wang2019beyond} introduce a general attack framework called mGAN-AI, which targets user-level privacy by identifying client identities. This framework was successfully applied to recover samples from specific users in the MNIST and AT\&T datasets.
He et al. \cite{he2019model} employ regularized maximum likelihood estimation in a white-box setting to achieve image restoration. In a black-box setting, they utilize inverse networks to identify the inverse mapping from the output to the input. They also offers feasible defense strategies to mitigate such attacks, including running fully-connected layers before releasing results, making client-side networks deeper, trusted execution on untrusted participants, differential privacy, and homomorphic encryption.
Zhang et al. \cite{zhang2020secret} leverage partial public information to reconstruct private data, extracting general knowledge from public datasets using GANs to regularize the inversion problem, and demonstrates that this attack remains highly effective even under constraints imposed on public datasets, highlighting that the defense strategy of differential privacy may not be sufficient to resist such attacks.
Zhu et al. \cite{zhu2020r} present a closed-form recursive procedure for recovering data from gradients of DNN through private recursive gradient attacks. Additionally, they propose rank analysis methods for estimating the risk of gradient attacks inherent in certain network architectures.
Kahla et al. \cite{kahla2022label} introduce the Boundary-Repelling Model Inversion (BREP-MI) technique, designed to invert private training data using predicted labels from the target model. This approach involves updating latent vectors with estimated gradients, enabling the GAN to reconstruct representative images of the target class from these latent vectors, and they also validates the effectiveness of this attack on various datasets and model architectures.
Han et al. \cite{han2023reinforcement} leverage confidence scores obtained from reinforcement learning-based GANs to provide rewards to agents. These rewards guide the agent's exploration of latent vectors for reconstructing private data, leading to the successful recovery of private information from target models across various datasets and models.
%\clearpage
\begin{table*}
\scriptsize
\caption{\textbf{Reviews of reconstruction attack libraries.} Taking the first line as an example, we utilized the original library [L.30] to conduct model attacks in a black-box environment. We used Python 2.7 with CPU, and the AT\&T Face dataset. However, the environment setup failed, and we were unable to debug the program to obtain results (e.g, the evaluation column is N/A).}
\label{tablera}
\begin{center}
\begin{tabular}{ccccccccccc}\hline
% Ref&Attack Type&White/Black-Box&\makecell[c]{Library Link\\
% (Appendix)}&\makecell[c]{Original Library}&Implementation&Python&GPU&Dataset&Evaluation&Issues\\ 
% \hline
\textbf{Ref.} & \textbf{Attack Type} & \textbf{White/Black-Box} & \textbf{\makecell[c]{Library Link \\ (Appendix)}} & \textbf{\makecell[c]{Original}} & \textbf{Impl.} & \textbf{Python} & \textbf{GPU} & \textbf{Dataset} & \textbf{Evaluation} & \textbf{Issues} \\
\hline\hline

\multirow{5}{*}{\makecell[c]{\cite{fredrikson2015model}}}  &\multirow{5}{*}{\makecell[c]{Model Attack}} &\multirow{5}{*}{\makecell[c]{Black-Box}}&\multirow{1}{*}{[L.30]} &\multirow{1}{*}{True} &\multirow{1}{*}{False} &\multirow{1}{*}{{2.7}} &\multirow{1}{*}{False} &{\makecell[c]{AT\&T \\Face}}&\makecell[c]{N/A}&{\makecell{Environment \\loading failed}}\\ 
\cline{4-11}

{}&{}&{}&{\makecell[c]{[L.31]}}& {\makecell[c]{False}}&{\makecell[c]{True}}&{\makecell[c]{3.10}}&{\makecell[c]{False}}&{\makecell[c]{AT\&T \\Face}}&{\makecell[c]{N/A}}
&{\makecell{Reproduced \\the function \\of reversing \\ the image}}\\
\hline

\cite{he2019model}&\makecell[c]{Data Attack}&{\makecell[c]{Black-Box}}&{\makecell[c]{[L.32]}}& {\makecell[c]{True}}&{\makecell[c]{True}}&{\makecell[c]{2.7}}&{\makecell[c]{True}}&{\makecell[c]{CIFAR-10}}&{\makecell[c]{Test Accuracy:  0.8729\\MSE: 0.1973}}
&{\makecell{Debugging \\solution}}\\
\hline

\multirow{7}{*}{\makecell[c]{\cite{erdougan2022unsplit}}}  &\multirow{7}{*}{\makecell[c]{Data Attack}} &\multirow{7}{*}{\makecell[c]{Black-Box}}&\multirow{7}{*}{[L.33]} &\multirow{7}{*}{True} &\multirow{7}{*}{True} &\multirow{7}{*}{{3.10}} &\multirow{7}{*}{True} &\multirow{1}{*}{MNIST}&\makecell[c]{Average MSE: 0.09\\
Clone Test Score: 0.05\\
Accuracy: 1.00
}&{\makecell{N/A}}\\
\cline{9-11}

{}&{}&{}&{}&{}&{}&{}&{}&{\makecell[c]{Fashion-\\MNIST}}&{\makecell[c]{Average MSE: 0.16\\
Clone Test Score:0.10\\
Accuracy: 1.00}}
&{\makecell{Debugging \\solution}}\\
\cline{9-11}

{}&{}&{}&{}&{}&{}&{}&{}&{\makecell[l]{CIFAR-10}}&{\makecell[c]{Average MSE: 0.10\\
Clone Test Score: 0.09\\
Accuracy: 1.00
}}&{\makecell{Debugging \\solution}}\\
\hline

\multirow{4}{*}{\makecell[c]{\cite{hitaj2017deep}}}  &\multirow{4}{*}{\makecell[c]{Data Attack}} &\multirow{4}{*}{\makecell[c]{Black-Box}}&\multirow{4}{*}{[L.34]} &\multirow{4}{*}{False} &\multirow{4}{*}{True} &\multirow{4}{*}{{3.10}} &\multirow{4}{*}{True} &\multirow{4}{*}{MNIST}&\makecell[c]{GAN:\\
Testing Accuracy: 0.95\\
Tseting Loss: 0.19}&{\makecell{N/A}}\\
\cline{10-11}

{}&{}&{}&{}&{}&{}&{}&{}&{}&{\makecell[c]{Federated\_GAN:\\
Testing Accuracy: 0.97\\
Testing Loss: 0.18}}
&{\makecell{Debugging \\solution}}\\
\hline

\cite{kahla2022label}&\makecell[c]{Model Attack}&{\makecell[c]{Black-Box}}&{\makecell[c]{[L.35]}}& {\makecell[c]{True}}&{\makecell[c]{True}}&{\makecell[c]{3.10}}&{\makecell[c]{True}}&{\makecell[c]{CelebA}}&{\makecell[c]{Total Accuracy: 0.90}}
&{\makecell{Debugging \\solution}}\\
\hline

\cite{han2023reinforcement}&\makecell[l]{Model Attack}&{\makecell[c]{Black-Box}}&{\makecell[c]{[L.36]}}& {\makecell[c]{True}}&{\makecell[c]{True}}&{\makecell[c]{3.10}}&{\makecell[c]{True}}&{\makecell[c]{CelebA}}&{\makecell[c]{Confidence score for the\\ target model : 0.9997\\ Accuracy : 1.000\\ Top-5 Accuracy : 1.000}}
&{\makecell{Debugging \\solution}}\\
\hline

\cite{balle2022reconstructing}&\makecell[c]{Model Attack}&{\makecell[c]{Black-Box}}&{\makecell[c]{[L.37]}}& {\makecell[c]{True}}&{\makecell[c]{True}}&{\makecell[c]{3.10}}&{\makecell[c]{False}}&{\makecell[c]{MNIST}}&{\makecell[c]{Training\_Loss: 0.02\\
Testing\_Loss: 0.06\\
Testing\_MAE: 0.04\\
Testing\_MSE: 0.02}}
&{\makecell{Debugging \\solution}}\\
\hline

\cite{zhu2020r}&\makecell[c]{Data Attack}&{\makecell[c]{Black-Box}}&{\makecell[c]{[L.38]}}& {\makecell[c]{True}}&{\makecell[c]{True}}&{\makecell[c]{3.10}}&{\makecell[c]{True}}&{\makecell[c]{CIFAR-10}}&{\makecell[c]{Lstsq Residual: 0.01062967\\max/min singular value:\\1.35e+00/1.40e-02}}
&{\makecell{N/A}}\\
\hline

\cite{yang2019neural}&\makecell[c]{Model Attack}&{\makecell[c]{Black-Box}}&{\makecell[c]{[L.39]}}& {\makecell[c]{False}}&{\makecell[c]{False}}&{\makecell[c]{3.10}}&{\makecell[c]{True}}&{\makecell[c]{CIFAR-10}}&{\makecell[c]{N/A}}
&{\makecell{Environment \\loading failed}}\\
\hline
 
\cite{song2017machine}&\makecell[c]{Model Attack}&{\makecell[c]{Black-Box}}&{\makecell[c]{[L.40]}}& {\makecell[c]{True}}&{\makecell[c]{False}}&{\makecell[c]{3.10}}&{\makecell[c]{True}}&{\makecell[c]{CelebA}}&{\makecell[c]{N/A}}
&{\makecell{Debugging \\ failed}}\\
\hline

\multirow{3}{*}{\makecell[c]{\cite{zhao2020idlg}}}  &\multirow{3}{*}{\makecell[c]{Data Attack}} &\multirow{3}{*}{\makecell[c]{Black-Box}}&\multirow{3}{*}{[L.41]} &\multirow{3}{*}{True} &\multirow{3}{*}{True} &\multirow{3}{*}{{3.10}} &\multirow{3}{*}{True} &\multirow{3}{*}{MNIST}&{\makecell[c]{Loss: 2.18e-08\\
MSE: 2.25e-08}
}&{\makecell{N/A}}\\
\cline{10-11}

{}&{}&{}&{}&{}&{}&{}&{}&{}&{\makecell[c]{Loss: 7.98e-06\\MSE: 4.79e-05}}&{\makecell{Debugging \\solution}}\\
\hline

{N/A}&{Data Attack}&{\makecell[c]{Black-Box}}&{\makecell[c]{[L.42]}}& {\makecell[c]{N/A}}&{\makecell[c]{True}}&{\makecell[c]{3.10}}&{\makecell[c]{True}}&UTKFace&{\makecell[c]{ Accuracy: 0.47}}
&{\makecell{Debugging\\failed}}\\
\hline

{N/A}&{Data Attack}&{\makecell[c]{Black-Box}}&{\makecell[c]{[L.43]}}& {\makecell[c]{N/A}}&{\makecell[c]{True}}&{\makecell[c]{3.10}}&{\makecell[c]{True}}&CelebA&{\makecell[c]{Testing Accuracy:0.55}}
&{\makecell{Debugging\\failed}}\\
\hline
 
\end{tabular}
\label{table3} 
\end{center}
\end{table*}

\subsection{\textbf{Property Inference Attacks}}\label{PIA}
\begin{figure}[!ht]
     \centering
     \includegraphics[width=0.50\textwidth]{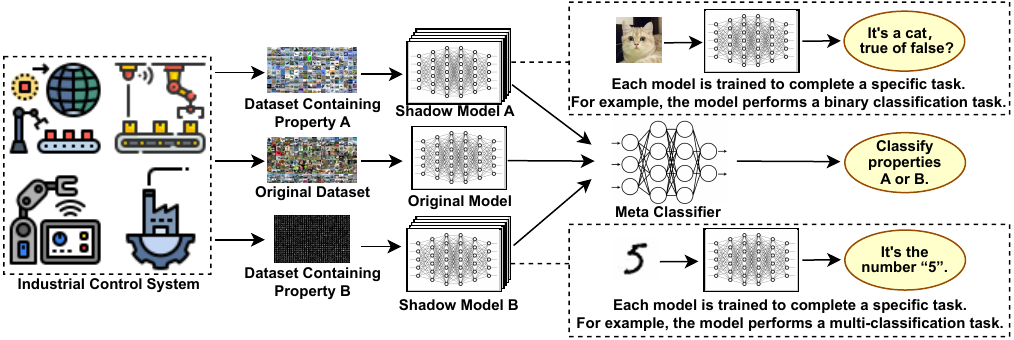}
     \caption{\textbf{Overview of property inference attacks in ML-based IoT system.}}
     \label{Fig4}
\end{figure}

Property inference attacks are based on the correlation between model output and known information. The attacker infers some sensitive properties of the input data by analyzing the relationship between the model output and known information. For example, in Fig. \ref{Fig4}. The attacker does not have access to the original dataset, but has access to the target model. If the attacker's target is property A, he needs to create a data set containing A, and then train multiple shadow models based on the created data set, with the same tasks as the target model, such as binary classification or multi-classification. Afterwards, the meta-classifier is trained on the shadow model in a supervised manner until the meta-classifier is able to infer the training data property A or B from the target model.

\hspace{-10pt}
Melis et al. \cite{melis2019exploiting} assume that the opponent has auxiliary data with property data points $({D^{adv}_{prop}})$ and non-property data points $({D^{adv}_{nonprop}}).$ The attacker uses the snapshot of the global model to generate aggregate updates of property data and updates of non-property data and labels the generated samples, thereby training a binary batch property classifier. In passive property inference attacks, the attacker observes updates and performs inference without changing the collaborative training content. Given a model snapshot $\theta_t,$ calculate the gradient $g_{prop}$; $({D^{adv}_{prop}})$ with property data and the gradient $g_{nonprop}$; $({D^{adv}_{nonprop}})$ without property data. After collecting enough label gradients, train the binary classifier $f_{prop.}$
\begin{align}
&g_{prop}=\bigtriangledown L({b^{adv}_{prop}};\theta_t)\notag\\ 
&G_{prop} \leftarrow G_{prop} \cup \lbrace g_{prop} \rbrace 
\end{align}
\begin{align}
&g_{prop}=\bigtriangledown L({b^{adv}_{nonprop}};\theta_t)\notag\\
&G_{nonprop} \leftarrow G_{nonprop} \cup \lbrace g_{nonprop} \rbrace
\end{align}
In an active property inference attack, the attacker extends a local copy of the collaboratively trained model with an enhanced attribute classifier connected to the last layer. The trained model performs well on the main task and recognizes batch attributes. For training data with each record having a primary label $y$ and an attribute label $p$, the joint loss of the model is calculated as:
\begin{align}
&L_{mt}=\alpha \cdot(x,y;\theta)+(1-\alpha) \cdot L(x,p;\theta)
\end{align}

%Generally, the process is divided into two phases. The first phase is when the attacker first inputs data and collects the output data of the target model.
%The second phase is that the attacker uses the inference algorithm to infer the input data's sensitive attributes by analyzing the model's output and known information. An attacker can obtain additional information about the input data through the inferred sensitive attributes. 

In this section, we show the experimental results in TABLE \ref{tablepia} and briefly introduce some early classic articles and current cutting-edge work on property inference attacks selected from \cite{ateniese2015hacking,ganju2018property,melis2019exploiting,song2019overlearning,xu2020subject,parisot2021property,malekzadeh2021honest,zhou2021property,zhang2021leakage,chase2021property,wang2022group,wang2022poisoning,mahloujifar2022property,hu2023prisampler,kerkouche2023client}.  (For some unintroduced papers, we also display them in TABLE \ref{tab}).

Ateniese et al. \cite{ateniese2015hacking} establish a new meta-classifier and by training the meta-classifier to crack other classifiers, obtain meaningful information about the training set of other classifiers. In contrast to previous work \cite{erman2007offline,tarca2007machine,nguyen2008survey,cohen2005packet,diao2009decision}, their approach focuses on statistics strictly related to the training samples used in the learning phase.
Ganju et al. \cite{ganju2018property} take advantage of the fact that FCNN is invariant under the arrangement of nodes in each layer to simplify the information extraction task, and put forward a feasible defense strategy for this attack, such as using node multiplicative transformations, adding noisy data to the training set and encoding arbitrary Information.
Melis et al. \cite{melis2019exploiting} use the collected data to generate aggregate updates and then train a binary attribute classifier and upload the joint loss function to determine whether the training data has the attribute. They also evaluate sharing fewer gradients \cite{shokri2015privacy}, dimensionality reduction, dropout \cite{salem2018ml}, and participant-level differential privacy \cite{mcmahan2017learning,geyer2017differentially}, find that they cannot effectively prevent attacks.
For over-learning, Song et al. \cite{song2019overlearning} developed a new transfer learning-based technique to evolve a benign training model into a malicious privacy-invading task model. A new learning-based review method is developed to transform the reviewed representation into a different form to extract relevant attribute information of interest.
To achieve agent-level privacy inference in the training phase, Xu et al. \cite{xu2020subject} propose CycleGAN to reconstruct the impact of data with specific attributes on the global model and implement gradient generation for inference attacks. Whether it is data points or some statistical features, the method can infer training data based on model parameters or intermediate outputs.
Zhang et al. \cite{zhang2021leakage} builds an attack vector based on a series of queries, performs hundreds of inference queries on the model, and jointly infers the attributes of a data set. They also point out that existing secure computing \cite{knott2021crypten} and differential privacy techniques \cite{dwork2014algorithmic,cormode2011personal} are either not directly applicable to certain scenarios or are costly.
Zhou et al. \cite{zhou2021property} generates generated samples by querying the target generator, then builds an attribute classifier to classify the generated samples, and finally classifies them according to the attributes of interest and then predicts the target attributes based on the output of the attribute classifier. In addition, they verify the feasibility of the attack on five property inference tasks, four datasets, and four victim GAN models.
To infer whether the training dataset is gender-balanced or not. Parisot et al. \cite{parisot2021property} train an attack model and input the target model's weights, and output the probability that the dataset used to train the target model has the attribute or does not have the attribute.
Malekzadeh et al. \cite{malekzadeh2021honest} proposes to train an honest but curious (HBC) classifier, and using the classifier can accurately predict its target attributes, and can also use the characteristics of its output to encode sensitive attributes to achieve attacks secretly. They also propose to add random noise to the output of the model, which is then shared to achieve defense.
Mahloujifar et al. \cite{mahloujifar2022property} point out that the adversary selects the attributes to attack, then uploads the input data in the poisoned distribution, and finally uses black-box queries (label queries only) on the trained model to determine the frequency of the selected attributes. Furthermore, they show that high attack accuracy can be achieved with a low degree of poisoning.
\begin{table*}
\caption{\textbf{Reviews of property inference attack libraries.} Taking the first line as an example, we utilized the original library [L.44] to conduct data attacks in a black-box environment. We used Python 2.7 with GPU, and the LFW dataset. After debugging, the results are shown in the evaluation.}
\label{tablepia}
\begin{center}\scriptsize
\begin{tabular}{ccccccccccc}\hline
% Ref&Attack Type&White/Black-Box&\makecell[c]{Library Link\\
% (Appendix)}&\makecell[c]{Original Library}&Implementation&Python&GPU&Dataset&Evaluation&Issues\\
% \hline
\textbf{Ref.} & \textbf{Attack Type} & \textbf{White/Black-Box} & \textbf{\makecell[c]{Library Link \\ (Appendix)}} & \textbf{\makecell[c]{Original}} & \textbf{Impl.} & \textbf{Python} & \textbf{GPU} & \textbf{Dataset} & \textbf{Evaluation} & \textbf{Issues} \\
\hline\hline

\cite{melis2019exploiting}&{Data Attack}&{\makecell[c]{Black-Box}}&{\makecell[c]{[L.44]}}& {\makecell[c]{True}}&{\makecell[c]{True}}&{\makecell[c]{2.7}}&{\makecell[c]{True}}&LFW&{\makecell[c]{Precision: 1.00\\
Recall: 0.99\\
F1-Score: 1.00\\
AUC: 1.00}}
&{\makecell{Debugging\\Solution}}\\
\hline 

\cite{malekzadeh2021honest}&\makecell[l]{Model Attack}&{\makecell[c]{Black-Box}}&{\makecell[c]{[L.45]}}& {\makecell[c]{True}}&{\makecell[c]{True}}&{\makecell[c]{3.10}}&{\makecell[c]{True}}&UTKFace&{\makecell[c]{Testing Accuracy: 0.82}}
&{\makecell{Debugging\\Solution}}\\
\hline 

\cite{suri2021formalizing}&{Data Attack}&{\makecell[c]{Black-Box}}&{\makecell[c]{[L.46]}}& {\makecell[c]{True}}&{\makecell[c]{False}}&{\makecell[c]{3.10}}&{\makecell[c]{True}}&CelebA&{\makecell[c]{N/A}}
&{\makecell{Debugging\\failed}}\\
\hline

\cite{song2019overlearning}&\makecell[l]{Model Attack}&{\makecell[c]{Black-Box}}&{\makecell[c]{[L.47]}}& {\makecell[c]{True}}&{\makecell[c]{False}}&{\makecell[c]{3.10}}&{\makecell[c]{True}}&UTKFace&{\makecell[c]{N/A}}
&{\makecell{Environment \\loading failed}}\\
\hline

\cite{hartmann2023distribution}&\makecell[l]{Data Attack}&{\makecell[c]{Black-box/\\White-Box}}&{\makecell[c]{[L.48]}}& {\makecell[c]{True}}&{\makecell[c]{True}}&{\makecell[c]{3.10}}&{\makecell[c]{True}}&N/A&{\makecell[c]{Output as picture}}
&{\makecell{Debugging\\Solution}}\\
\hline

\cite{suri2023dissecting}&\makecell[l]{Model Attack}&{\makecell[c]{Black-Box}}&{\makecell[c]{[L.49]}}& {\makecell[c]{True}}&{\makecell[c]{False}}&{\makecell[c]{3.10}}&{\makecell[c]{True}}&CelebA&{\makecell[c]{N/A}}
&{\makecell{Environment \\loading failed}}\\
\hline

\cite{parisot2021property}&\makecell[l]{Data Attack}&{\makecell[c]{Black-Box}}&{\makecell[c]{[L.50]}}& {\makecell[c]{True}}&{\makecell[c]{False}}&{\makecell[c]{3.10}}&{\makecell[c]{True}}&CelebA&{\makecell[c]{N/A}}
&{\makecell{Environment \\loading failed}}\\
\hline

\cite{pasquini2021unleashing}&\makecell[l]{Model Attack}&{\makecell[c]{N/A}}&{\makecell[c]{[L.51]}}& {\makecell[c]{True}}&{\makecell[c]{False}}&{\makecell[c]{3.10}}&{\makecell[c]{True}}&MNIST&{\makecell[c]{Validation: 0.0193}}
&{\makecell{N/A}}\\
\hline

%\caption{Table 4.Review the library of Property inference attacks}
%\label{table}
\end{tabular}
\label{table4} 
\end{center}
\end{table*}

\subsection{\textbf{Model Extraction Attacks}}\label{MEA}
\begin{figure}[!ht]
     \centering
     \includegraphics[width=0.50\textwidth]{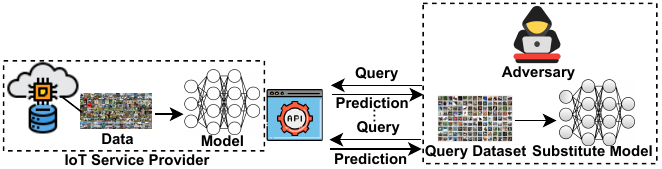}
     \caption{\textbf{Overview of model extraction attacks in ML-based IoT system.}}
     \label{Fig5}
\end{figure}

Model extraction attacks are based on training data and black-box model input and output. By observing the input and output of the black box model, the attacker infers the characteristics and rules of the input data and output results so as to train the approximate model to simulate the behavior of the black box model. The process is shown in Fig. \ref{Fig5}.

Tramèr et al. \cite{tramer2016stealing} model the adversary as a stochastic algorithm. The adversary's goal is to use as few query inputs $f$ as possible in order to efficiently compute an approximation that closely matches the input $f$. “Tight matching” is formalized using two different error measures.

This is the average error on the test set $D$, a lower test error means that for inputs distributed like the training data samples, $\hat{f}$ matches $f$ well.
\begin{flalign}
&R_{test}=(f,\hat{f}) = \sum\nolimits_{{(x,y)}\in D }d(f(x),\hat{f}(x))/|D| 
\end{flalign}

For a set $U$ of vectors chosen uniformly in $\chi$, In the following formula, $R_{unif}$ estimates the proportion of the complete feature space where $f$ and $\hat{f}$ are inconsistent:
\begin{flalign}
&R_{unif}=(f,\hat{f}) = \sum\nolimits_{{x}\in U }d(f(x),\hat{f}(x))/|U| 
\end{flalign}

%Generally, the process is divided into two phases. The first phase is that the attacker inserts malicious samples into the acquired training dataset.
%The second phase is that the attacker provides the tampered training data set to the model for training process. Models trained with tampered training data lead to learning wrong patterns or making wrong decisions during training.
%The overflow is shown in Fig. \ref{Fig5}.
In this section, we show the experimental results in TABLE \ref{tableema} and briefly introduce some early classic articles and current cutting-edge work on model extraction attacks selected from \cite{tramer2016stealing,wang2018stealing,correia2018copycat,hong2018security,oh2019towards,orekondy2019knockoff,yoshida2019model,batina2019csi,juuti2019prada,reith2019efficiently,krishna2019thieves,jagielski2020high,carlini2020cryptanalytic,chandrasekaran2020exploring,aivodji2020model,fukuoka2020model,zhu2021hermes,miura2021megex,hu2021model,gong2021inversenet,carlini2021extracting,zhang2021seat,truong2021data,wu2022model,dziedzic2022dataset,dziedzic2022increasing,liu2022stolenencoder,chen2023d,chang2023rethink}. (For some unintroduced papers, we also display them in TABLE \ref{tab}).

Tramèr et al. \cite{tramer2016stealing} describe an approach to access a target model as a black box, input data, observe the model's output, and construct a training dataset based on these outputs, and extract parameters or structural information from the model using techniques like querying the prediction API and solving equations. They discuss several defense strategies, including rounding confidences \cite{fredrikson2015model}, differential privacy \cite{dwork2006differential,li2013membership}, and ensemble methods.
Wang et al. \cite{wang2018stealing} propose a general attack framework for stealing hyperparameters. They compute the gradient of the objective function at the model parameters and use linear least squares to estimate the hyperparameters. Furthermore, they demonstrate that $L$2 regularization can be more effective in defending against this attack. The defensive effects of cross-entropy and square hinge loss are similar.
Correia-Silva et al. \cite{correia2018copycat} query the target network with random unlabeled data, creating a fake dataset using input image and output label pairs to imitate the target network's knowledge, and train an imitation network on this fake dataset to achieve similar performance to the target network. They experimentally show that a CNN can be created by treating the target network query as a black box for random unlabeled data. In contrast to previous work \cite{oh2019towards,papernot2017practical,tramer2016stealing}, Orekondy et al. \cite{orekondy2019knockoff} impose constraints on the adversary's capabilities. The adversary first queries a black-box model for a set of input images to obtain predictions and then uses these predictions to train similar images. Although this method makes fewer black-box assumptions, it demonstrates the power of the attack.
Juuti et al. \cite{juuti2019prada} access an independent target model as a black box, obtain the output responses for a corresponding query set, initialize the model's hyperparameters, synthesize samples, and iteratively train the model. In addition, They discuss feasible defense methods, such as modifying the model's predictions \cite{tramer2016stealing,lee2019defending}, and detecting adversarial examples \cite{grosse2017statistical,meng2017magnet}.
Oh et al. \cite{oh2019towards} introduce a meta-model called kennen, which can infer model attributes from static query inputs and actively optimize query inputs to extract more information. They also demonstrate that even with just 100 single-label outputs, a wealth of information about black boxes can be revealed, and black-box access can be predicted with high accuracy even as the metamodel's performance degrades.
Batina et al. \cite{batina2019csi} base their approach on reverse engineering of neural networks using side-channel analysis techniques. They collect measured data corresponding to the selected network implementation and recover key parameters. Additionally, they propose several defense strategies, such as shuffling \cite{veyrat2012shuffling}, masking, and constant-time implementations of exponentiation \cite{al2008comparative}.
Jagielski et al. \cite{jagielski2020high} develop a learning-based attack aimed at extracting high-precision models, using the victim to supervise the training of the extracted model. Using a hybrid approach, a learning-based approach is utilized to recover from numerical instability errors while performing a functional equivalence extraction attack. It is worth noting that this method does not require training and can directly extract the model's weights.
Carlini et al. \cite{carlini2021extracting} describe an attack that guides the model to generate diverse output results through different sampling strategies to discover the samples memorized by the model, and use a specified measurement method to measure the model's familiarity with the samples and determine which samples are memorized. They also propose mitigation strategies, such as training with differential privacy \cite{abadi2016deep}, curating the training data to limit the impact of memorization on downstream applications, and auditing ML models for memorization.
Wu et al. \cite{wu2022model} propose a model extraction attack on Graph Neural Networks (GNNs), and target a private GNN model, constructing a replica model with similar functionality by querying the sequence of nodes obtained from the target model. They demonstrate that this attack method can effectively extract duplicate models resembling the target model and achieve highly accurate prediction results.
%\clearpage
\begin{table*}
\caption{\textbf{Reviews of model extraction attack libraries.} Taking the first line as an example, we utilized the original library [L.52] to conduct model attacks in a black-box environment. We used Python 2.7 with CPU, and the MNIST dataset. After debugging, the results are shown in the evaluation.}
\label{tableema}
\begin{center}\scriptsize
\begin{tabular}{ccccccccccc}\hline
% Ref&Attack Type&White/Black-Box&\makecell[c]{Library Link\\
% (Appendix)}&{\makecell[c]{Original Library}}&Implementation&Python&GPU&Dataset&Evaluation&Issues\\ 
% \hline
\textbf{Ref.} & \textbf{Attack Type} & \textbf{White/Black-Box} & \textbf{\makecell[c]{Library Link \\ (Appendix)}} & \textbf{\makecell[c]{Original}} & \textbf{Impl.} & \textbf{Python} & \textbf{GPU} & \textbf{Dataset} & \textbf{Evaluation} & \textbf{Issues} \\
\hline\hline

\multirow{27}{*}{\makecell[c]{\cite{tramer2016stealing}}}  &\multirow{27}{*}{\makecell[c]{Model Attack}} &\multirow{27}{*}{\makecell[c]{Black-Box}}&\multirow{27}{*}{[L.52]} &\multirow{27}{*}{True} &\multirow{27}{*}{True} &\multirow{27}{*}{2.7} &\multirow{27}{*}{False} &\multirow{1}{*}{MNIST}&{\makecell[c]{Training Accuracy: 0.96}}
&{\makecell{Debugging \\solution}}\\
\cline{9-11}
 
{}&{}&{}&{}&{}&{}&{}&{}&{Circles}&{\makecell[c]{Training Accuracy: 0.44}}&{\makecell{Debugging \\solution}}\\
\cline{9-11}

{}&{}&{}&{}&{}&{}&{}&{}&{Blobs}&{\makecell[c]{Training Accuracy: 1.00}}&{\makecell{Debugging \\solution}}\\
\cline{9-11}

{}&{}&{}&{}&{}&{}&{}&{}&{Adult}&{\makecell[c]{Training Accuracy: 0.88}}&{\makecell{Debugging \\solution}}\\
\cline{9-11}

{}&{}&{}&{}&{}&{}&{}&{}&{Moons}&{\makecell[c]{Training Accuracy: 0.86}}&{\makecell{Debugging \\solution}}\\
\cline{9-11}

{}&{}&{}&{}&{}&{}&{}&{}&{Adult\_b}&{\makecell[c]{Training Accuracy: 0.85}}&{\makecell{Debugging \\solution}}\\
\cline{9-11}

{}&{}&{}&{}&{}&{}&{}&{}&{Class5}&{\makecell[c]{Training Accuracy: 0.57}}&{\makecell{Debugging \\solution}}\\
\cline{9-11}

{}&{}&{}&{}&{}&{}&{}&{}&{Digits}&{\makecell[c]{Training Accuracy: 0.98}}&{\makecell{Debugging \\solution}}\\
\cline{9-11}

{}&{}&{}&{}&{}&{}&{}&{}&{Face}&{\makecell[c]{Training Accuracy: 0.96}}&{\makecell{Debugging \\solution}}\\
\cline{9-11}

{}&{}&{}&{}&{}&{}&{}&{}&{Steak}&{\makecell[c]{Training Accuracy: 0.44}}&{\makecell{Debugging \\solution}}\\
\cline{9-11}

{}&{}&{}&{}&{}&{}&{}&{}&{Gss}&{\makecell[c]{Training Accuracy: 0.64}}&{\makecell{Debugging \\solution}}\\
\cline{9-11}

{}&{}&{}&{}&{}&{}&{}&{}&{Iris}&{\makecell[c]{Training Accuracy: 0.97}}&{\makecell{Debugging \\solution}}\\
\cline{9-11}

{}&{}&{}&{}&{}&{}&{}&{}&{Diabetes}&{\makecell[c]{Training Accuracy: 0.79}}&{\makecell{Debugging \\solution}}\\
\cline{9-11}

{}&{}&{}&{}&{}&{}&{}&{}&{Mushrooms}&{\makecell[c]{Training Accuracy: 1.00}}&{\makecell{Debugging \\solution}}\\
\cline{9-11}

{}&{}&{}&{}&{}&{}&{}&{}&{Cancer}&{\makecell[c]{Training accuracy: 0.96}}&{\makecell{Debugging \\solution}}\\
\hline

\cite{wu2022model}&\makecell[l]{Model Attack}&{\makecell[c]{Black-Box/\\White-Box}}    &{\makecell[c]{[L.53]}}& {\makecell[c]{True}}&{\makecell[c]{True}}&{\makecell[c]{2.7}}&{\makecell[c]{False}}&{\makecell[c]{Cora}}&{\makecell[c]{Accuracy:0.13}}
&{\makecell{Debugging \\solution}}\\
\hline

\cite{carlini2021extracting}&\makecell[l]{Data Attack}&{\makecell[c]{Black-Box}}    &{\makecell[c]{[L.54]}}& {\makecell[c]{True}}&{\makecell[c]{True}}&{\makecell[c]{3.10}}&{\makecell[c]{True}}&{\makecell[c]{Common \\Crawl}}&{\makecell[c]{Top sample by ratio\\ of Zlib entropy\\ and XL perplexity:\\PPL-XL: 12.805\\Zlib: 248.000\\Score: 97.260}}
&{\makecell{Debugging \\solution}}\\
\hline

\cite{barbalau2020black}&\makecell[l]{Data Attack}&{\makecell[c]{Black-Box}}    &{\makecell[c]{[L.55]}}& {\makecell[c]{True}}&{\makecell[c]{True}}&{\makecell[c]{3.10}}&{\makecell[c]{True}}&{\makecell[c]{CIFAR-10}}&{\makecell[c]{Accuracy: 0.83}}
&{\makecell{Debugging \\solution}}\\
\hline

\cite{carlini2020cryptanalytic}&\makecell[l]{Model Attack}&{\makecell[c]{Black-Box}}    &{\makecell[c]{[L.56]}}& {\makecell[c]{True}}&{\makecell[c]{True}}&{\makecell[c]{3.7}}&{\makecell[c]{False}}&{\makecell[c]{N/A \\(By random\\ training data)
}}&{\makecell[c]{Maximum logit \\loss on the\\ unit sphere: 4.47e-12}}
&{\makecell{Debugging \\solution}}\\
\hline

\cite{oh2019towards}&\makecell[l]{Data Attack}&{\makecell[c]{Black-Box}}    &{\makecell[c]{[L.57]}}& {\makecell[c]{True}}&{\makecell[c]{True}}&{\makecell[c]{2.7}}&{\makecell[c]{True}}&{\makecell[c]{MNIST}}&{\makecell[c]{Average Loss: 0.0374\\ Accuracy: 0.9894}}
&{\makecell{Debugging \\solution}}\\
\hline

\cite{dziedzic2022dataset}&\makecell[l]{Data Attack}&{\makecell[c]{Black-Box}}    &{\makecell[c]{[L.58]}}& {\makecell[c]{False}}&{\makecell[c]{False}}&{\makecell[c]{3.10}}&{\makecell[c]{True}}&{CIFAR-10}&{\makecell[c]{N/A}}
&{\makecell{Debugging \\failed}}\\
\hline

\cite{correia2018copycat}&\makecell[l]{Data Attack}&{\makecell[c]{Black-Box}}    &{\makecell[c]{[L.59]}}& {\makecell[c]{False}}&{\makecell[c]{False}}&{\makecell[c]{3.10}}&{\makecell[c]{True}}&{CIFAR-10}&{\makecell[c]{N/A}}
&{\makecell{Debugging \\failed}}\\
\hline

\end{tabular}
\label{table5} 
\end{center}
\end{table*}

\subsection{\textbf{Poisoning Attacks}}\label{PA}
Poisoning attacks are based on tampering with the training data set or directly acting on the model to affect the training process of the machine learning system, guiding the model to learn wrong patterns or make wrong decisions. The process is shown in Fig. \ref{Fig5}.

Let $x$ denote the function that propagates the input $x$ through the network to the penultimate layer (before the softmax layer). Due to the high complexity and nonlinearity of $f$, it is possible to find a sample that “collides” with the target in the feature space and is close to the base sample b of the input space through the following formula $b.$
\begin{flalign}
&p = \underset{x}{argmin}\text{ }{||f(x)-f(t)||^{2}_{2}}+\beta\text{ }{||x-b||^{2}_{2}}
\end{flalign}

In order to prevent the separation of poisoned samples and targets during the training process, Shafahi et al. \cite{shafahi2018poison} use a simple but effective trick: adding a low-opacity watermark of the target sample to the poisoned samples, thereby allowing certain inseparable features to overlap while maintaining visual different. By weighted combination of the base image $b$ and the target image $t$, a basic watermark image with target opacity $\gamma$ is formed:
\begin{flalign}
&b \leftarrow \gamma \cdot t+(1-\gamma) \cdot b
\end{flalign}

%The attacker first obtains the training dataset and inserts malicious samples into it. Then provide the tampered training data set to the model for the training process. Models trained with tampered training data lead to learning wrong patterns or making wrong decisions during training. 
%The overflow is shown in Fig. {\ref{Fig6}}.
In this section, we show the experimental results in TABLE \ref{tablepa} and briefly introduce some early classic articles and current cutting-edge work on poisoning attacks selected from \cite{biggio2012poisoning,alfeld2016data,yang2017generative,chen2017targeted,shafahi2018poison,fang2018poisoning,jagielski2018manipulating,bhagoji2019analyzing,tolpegin2020data,kurita2020weight,zhang2021data,fang2020local,huang2021data,shejwalkar2021manipulating,rong2022fedrecattack,yerlikaya2022data,cao2022mpaf,yang2023data,gupta2023novel,cao2023fedrecover}. (For some unintroduced papers, we also display them in TABLE 
 \ref{tab}).

\begin{figure}[!ht]
     \centering
     \includegraphics[width=0.50\textwidth]{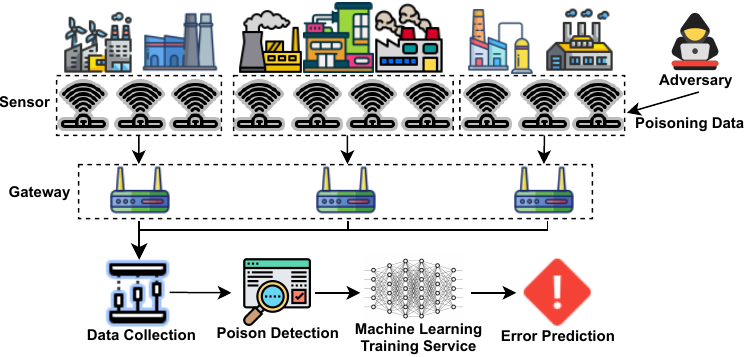}
     \caption{\textbf{Overview of poisoning attacks in ML-based IoT system.}}
     \label{Fig6}
\end{figure}
\\
Biggio et al. \cite{biggio2012poisoning} employs a gradient rising strategy, allowing adversaries to predict changes in the decision function of Support Vector Machines (SVM) caused by malicious input. They use this capability to construct malicious data and demonstrate that the gradient ascent process, coupled with accurate recognition, can increase the test error of the classifier.
Alfeld et al. \cite{alfeld2016data} propose a general mathematical framework and encode real-world settings into this framework. This enables the optimization problem to be solved using standard convex optimization methods. They demonstrate the attack's effectiveness on synthetic and real-world time-series data, comparing it to random and greedy baselines.
Shafahi et al. \cite{shafahi2018poison} introduce an optimization-based method for crafting poison data. In transfer learning scenarios, a single poison image can control the classifier's behavior. They also propose a "watermarking" strategy for full end-to-end training, using multiple poisoning training instances to enhance the reliability of poisoning attacks.
Jagielski et al. \cite{jagielski2018manipulating} adapt a model initially designed for poisoning attacks on classification problems to regression problems. They design a fast statistical attack that requires limited information to be effective. Additionally, they propose a defense algorithm called TRIM \cite{jagielski2018manipulating}, which outperforms traditional robust statistics \cite{fischler1981random,huber1992robust}.
An alternate minimization strategy is presented by Bhagoji et al. \cite{bhagoji2019analyzing}, involving the optimization of training loss and the adversarial objective alternately. This approach utilizes a set of explainability techniques to conduct model poisoning attacks while maintaining stealth. They demonstrate that model poisoning attacks can be carried out stealthily despite strict adversary constraints.
Fang et al. \cite{fang2020local} propose a local model poisoning attack for Byzantine inverse federated learning. The attack manipulates local model parameters on compromised working devices during the learning process. Furthermore, they introduce RONI \cite{barreno2010security} and TRIM \cite{jagielski2018manipulating} as defenses against poisoning attacks.
Tolpegin et al. \cite{tolpegin2020data} exploit a malicious subset of actors in the system to poison the global model by sending model updates originating from mislabeled data. Additionally, they propose a defense mechanism that enables aggregators to identify malicious actors.
Huang et al. \cite{huang2021data} aim to manipulate recommender systems by recommending target items to users. They construct a poison model, generate fake users, and inject them into the recommendation system to influence recommendations to regular users. They also validate the effectiveness and competitiveness of the attack by deploying detectors.
Zhang et al. \cite{zhang2021data} introduce RAPU-G, which leverages user and item data within a two-tier optimization framework to generate fake user and item interactions. Furthermore, they employ RAPU-R, in conjunction with context-specific heuristic rules, to handle data incompleteness and perturbation.
Rong et al. \cite{rong2022fedrecattack} propose FedRecAttack to increase the exposure rate of the target project. By approximating user feature vectors using public interactions, attackers can generate corresponding toxic gradients and manipulate malicious users to upload these toxic gradients.
\begin{table*}
\caption{\textbf{Reviews of poisoning attack libraries.} Taking the first line as an example, we utilized the original library [L.60] to conduct data attacks. We used Python 3.7 with CPU and the Fashion-MNIST dataset. After debugging, the results are shown in the evaluation.}
\label{tablepa}
\begin{center}\scriptsize
\begin{tabular}{cccccccccc}\hline
\textbf{Ref.} & \textbf{Attack Type} & \textbf{\makecell[c]{Library Link \\ (Appendix)}} & \textbf{\makecell[c]{Original}} & \textbf{Impl.} & \textbf{Python} & \textbf{GPU} & \textbf{Dataset} & \textbf{Evaluation} & \textbf{Issues} \\
\hline\hline
\cite{bhagoji2019analyzing}&Model Attack&[L.60]&True&True&3.7&False&\makecell[c]{Fashion-\\MNIST}&\makecell[c]{Accuracy: 0.9919}&\makecell[c]{Debugging \\solution}\\ 
\hline
\cite{tolpegin2020data}&Data Attack&[L.61]&True&True&3.7&False&\makecell[c]{Fashion-\\MNIST}&\makecell[c]{Precison: 79\\
Recall: 0.96\\
F1-Score: 0.85\\
Accuracy: 0.85}&\makecell[c]{N/A}\\ 
\hline
\cite{rong2022fedrecattack}&Model Attack&[L.62]&True&True&3.10&True&\makecell[c]{MovieLens}&\makecell[c]{Loss = 21.78348}&\makecell[c]{Debugging \\solution}\\ 
\hline
\multirow{5}{*}{\cite{zhang2021data}}  &\multirow{5}{*}{\makecell[c]{Data Attack}} &\multirow{5}{*}{\makecell[c]{[L.63]}} &\multirow{5}{*}{\makecell[c]{True}} &\multirow{5}{*}{\makecell[c]{True}} &\multirow{5}{*}{\makecell[c]{3.10}} &\multirow{5}{*}{\makecell[c]{True}} &\multirow{5}{*}{\makecell[c]{MovieLens}} &\makecell[c]{ RAPU\_R:\\
Recall: 0.1825\\
Precision: 0.0024\\ 
MRR: 0.0099
}&{\makecell[c]{Debugging \\solution}} \\ 
\cline{9-10}
   & & & & & & & &{\makecell[c]{ RAPU\_R:\\
Recall: 0.1825\\
Precision: 0.0024\\ 
MRR: 0.0099
}} & {\makecell[c]{Debugging \\solution}}\\ 
\hline
{N/A}&Data Attack&[L.64]&N/A&True&3.7&False&CIFAR-10&\makecell[c]{True\_Norm\_Accuracy: 1.00\\
True\_Poisoned\_Accuracy: 0.98\\
Norm\_Poisoned\_Accuracy: 0.98}&{\makecell{Debugging \\solution}}\\ 
\hline
{N/A}&Model Attack&[L.65]&N/A&True&3.10&True&Adult&\makecell[c]{Training Accuracy: 0.7287\\
Testing Accuracy: 0.7299}&{\makecell{Debugging \\solution}}\\ 
\hline
\end{tabular}
\label{table6} 
\end{center}
\end{table*}

%\begin{figure}
    %\tabcolsep=1.0\linewidth
%    \centering
%    \includegraphics[width=0.5\textwidth]{Picture/Thermal map.pdf}
%    \caption{Summary of papers on six types of machine learning attacks. We summarized previous work representative of each type of attack according to year (the closer to the center of the circle, the earlier the representative year) and recent work (the further away from the center of the circle, the later the year represented).}
    %\label{Fig10}
%\end{figure}

\section{\textbf{Discussion}}
In the rapidly expanding landscape of ML-based IoT systems, addressing security challenges becomes crucial. The taxonomy of attacks has grown to six categories: membership inference, adversarial, reconstruction, property inference, model extraction, and poisoning attacks. Protecting data security and privacy requires collaborative efforts, extending beyond technical and academic circles to society. The integration of IoT adds complexity, demanding advanced cryptographic techniques, secure communication protocols, and anomaly detection. Expect increased research for innovative solutions to enhance security and resilience. Privacy-preserving methods, robust models, and strict security protocols are essential. Public, industry, and stakeholders must uphold ethical standards and legal regulations to ensure privacy and data protection. Collaborative synergies, continuous scholarly inquiry, and innovative breakthroughs are imperative for leveraging ML-based IoT systems for societal advancement and well-being.

\begin{table}[!ht]
%\tabcolsep=0.001\linewidth
\centering
\caption{\textbf{Summary of papers on six types of attacks on ML-based IoT systems.} We summarized previous work representative of each type of attack according to year. (The heat map above illustrates the quantities of classical papers and projects. Each row corresponds to the inheritance relationships of respective attacks over the years. The table below provides detailed references).}
\includegraphics[width=0.5\textwidth]{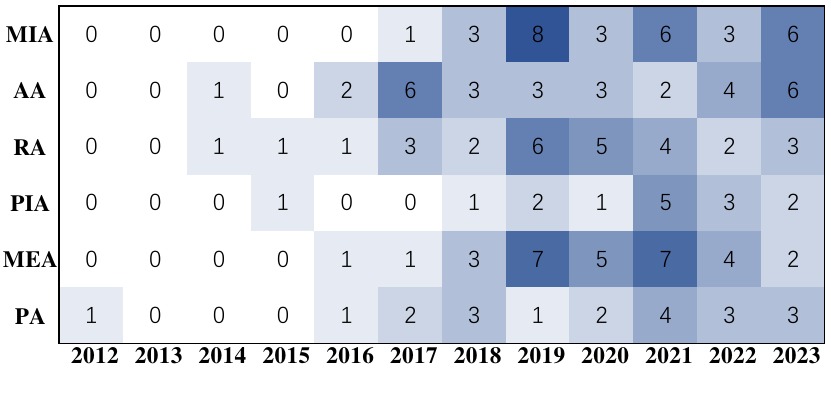}
\renewcommand{\arraystretch}{1.5}
\label{tab}
\resizebox{0.5\textwidth}{!}{
\begin{tabular}{ccccccc}
\hline
\diagbox{\textbf{Year}}{\textbf{Type}}& \textbf{MIA} & \textbf{AA} & \textbf{RA} & \textbf{PIA} & \textbf{MEA} & \textbf{PA} \\
\hline
\textbf{2012} & N/A & N/A & N/A & N/A & N/A &  \cite{biggio2012poisoning}\\
\hline
\textbf{2013} & N/A & N/A & N/A & N/A & N/A &  N/A\\
\hline
\textbf{2014} & N/A &\cite{goodfellow2014explaining} & \cite{fredrikson2014privacy} & N/A & N/A &  N/A\\
\hline
\textbf{2015} & N/A & N/A & \cite{fredrikson2015model} & \cite{ateniese2015hacking} & N/A &  N/A\\
\hline
\textbf{2016} & N/A & \cite{moosavi2016deepfool,kurakin2016adversarial} & \cite{wu2016methodology} & N/A & \cite{tramer2016stealing} & \cite{alfeld2016data}\\
\hline
\textbf{2017} & \cite{shokri2017membership} & \makecell[c]{\cite{madry2017towards,huang2017adversarial,carlini2017towards,chen2017zoo,papernot2017practical,brendel2017decision}} & \cite{hitaj2017deep,song2017machine,hidano2017model} & N/A & \cite{hitaj2017deep} &  \cite{yang2017generative,chen2017targeted}\\
\hline
\textbf{2018} & \cite{long2018understanding,yeom2018privacy,rahman2018membership} & \cite{zugner2018adversarial,dong2018boosting,ilyas2018black} & \cite{sannai2018reconstruction,lacharite2018improved} & \cite{ganju2018property} & \cite{wang2018stealing,correia2018copycat,hong2018security} & \cite{shafahi2018poison,fang2018poisoning,jagielski2018manipulating}\\
\hline
\textbf{2019} &\makecell[c]{\cite{song2019privacy,sablayrolles2019white,wang2019beyond,yaghini2019disparate,truex2019demystifying,irolla2019demystifying,song2019membership,truex2019effects}}
& \cite{bojchevski2019adversarial,guo2019simple,su2019one} & \makecell[c]{\cite{carlini2019secret,he2019model,yang2019neural,zhu2019deep,aivodji2019gamin,oh2019exploring}} & \cite{melis2019exploiting,song2019overlearning} & \makecell[c]{\cite{oh2019towards,orekondy2019knockoff,yoshida2019model,batina2019csi,juuti2019prada,reith2019efficiently,krishna2019thieves}} &  \cite{bhagoji2019analyzing}\\
\hline
\textbf{2020} & \cite{long2020pragmatic,leino2020stolen,humphries2020differentially}  & \cite{chen2020hopskipjumpattack,zugner2020adversarial,bin2020adversarial} & \makecell[c]{\cite{salem2020updates,zhang2020secret,xu2020midas,goldsteen2020reducing,zhu2020r}} & \cite{xu2020subject} & \makecell[c]{\cite{jagielski2020high,carlini2020cryptanalytic,chandrasekaran2020exploring,aivodji2020model,fukuoka2020model}} &  \cite{tolpegin2020data,kurita2020weight}\\
\hline
\textbf{2021} & \makecell[c]{\cite{chen2021machine,hui2021practical,song2021systematic,shokri2021privacy,hu2021source,choquette2021label}} &\cite{wang2021admix,rony2021augmented}& \cite{wang2021variational,zhao2021exploiting,dong2021privacy,kahla2022label} & \makecell[c]{\cite{parisot2021property,malekzadeh2021honest,zhou2021property,zhang2021leakage,chase2021property}} & \makecell[c]{\cite{zhu2021hermes,miura2021megex,hu2021model,gong2021inversenet,carlini2021extracting,zhang2021seat,truong2021data}} & \cite{zhang2021data,fang2020local,huang2021data,shejwalkar2021manipulating}\\
\hline
\textbf{2022} & \cite{liu2022membership,carlini2022membership,ye2022enhanced} & \cite{amich2022eg,jia2022adversarial,li2022subspace,xiong2022stochastic} & \cite{struppek2022plug,balle2022reconstructing} & \cite{wang2022group,wang2022poisoning,mahloujifar2022property} & \cite{wu2022model,dziedzic2022increasing,dziedzic2022dataset,liu2022stolenencoder} &  \cite{rong2022fedrecattack,yerlikaya2022data,cao2022mpaf}\\
\hline
\textbf{2023} & \makecell[c]{\cite{van2023membership,yuan2023interaction,matsumoto2023membership,bertran2023scalable,zhang2023mida,hu2023membership}} & \makecell[c]{\cite{song2023robust,kim2023demystifying,feng2023dynamic,wang2023towards,chen2023effective,huang2023t}} & \cite{han2023reinforcement,annamalai2023linear,long2023membership} & \cite{hu2023prisampler,kerkouche2023client} & \cite{chen2023d,chang2023rethink} & \cite{yang2023data,gupta2023novel,cao2023fedrecover}\\
\hline
\end{tabular}}
\end{table}

\subsection{\textbf{Experimental Configurations}}
In this study, we adopt two different experimental configurations: one part is performed on GPU, and the other part is performed on CPU (The environments corresponding to the specific experiments have been shown in TABLE \ref{table1} to TABLE \ref{table6}). We used the Ubuntu 20.04 LTS operating system in the experiment to maintain the consistency of the experimental environment and fully explore attacks under different hardware environments.
\subsubsection{\textbf{GPU Experimental Configuration}}
In the GPU experiment part, we used NVIDIA Tesla P40 GPU, paired with Intel Xeon E5-2683 v4 processor, equipped with 24GB GPU memory, 32GB RAM, and 1TB hard drive.

\subsubsection{\textbf{CPU Experimental Configuration}}
In the CPU experiment part, we selected the Intel Core i5-7500 processor, which has 4 cores, 8GB of RAM, and 1TB hard drive.

\subsection{\textbf{Open Source Project Evaluation}}
We conducted an extensive review of open-source projects to assess the performance and usability of our attacks. Our evaluation focused on two main attack categories: data attacks and model attacks. We analyzed these open-source projects for both white-box and black-box characteristics. To provide a comprehensive perspective, we categorized the projects into two groups: original libraries (open-sourced by the authors of this article) and unoriginal libraries (contributed by individuals or entities). This classification allowed us to gauge the overall usability of these open-source resources. Our evaluation process also included tracking the implementation of these projects to determine their reproducibility and practicality. We used various Python versions, including Python 2.7, Python 3.7, and Python 3.10, to ensure compatibility with different library environments. We noted whether these open-source projects supported GPUs, taking hardware dependencies into account. Additionally, we documented the dataset used in these open-source projects and provided comprehensive evaluation results, including performance metrics that highlight the strengths and weaknesses of each project. Furthermore, we identified common issues encountered during the testing of these projects and offered corresponding solutions or categorized the types of problems faced. This document serves as a valuable resource for other researchers, offering insights into the usability and limitations of these projects.

\subsection{\textbf{Adaptability to Different Hardware Configurations and Python Versions}}
Throughout the experiment, we faced several challenges, including variations in hardware configurations, Python versions, and parameter settings within open-source projects. We took measures to resolve these challenges and ensure the experiment's validity.

\subsubsection{\textbf{Adaptability to Different Hardware Configurations}}
We assessed adaptability to various hardware configurations by conducting tests in both GPU and CPU environments.

\subsubsection{\textbf{Python Version Differences}}
Given that different open-source projects may demand varying Python versions, we selected the appropriate Python version for compatibility based on each project's specific requirements.

\subsubsection{\textbf{Parameter Setting Issues}}
Some open-source projects had excessively large parameters set by their authors, resulting in lengthy execution times. To address this, we fine-tuned the parameters to ensure that experiments ran within acceptable time frames, considering our available computing resources.

\subsection{\textbf{Suggestions and Explorations}}
\subsubsection{\textbf{Recommendation on Library Usage}}
Based on the experimental results we have presented, we recommend that researchers prioritize the use of original libraries, as unoriginal libraries possess certain limitations, such as cases where individuals or entities utilized frameworks different from those of the authors \cite{shokri2017membership} (original library [L.1] and unoriginal libraries [L.2], [L.3]), faced changes in attack scenarios \cite{shokri2017membership} (original library [L.1] and unoriginal libraries [L.2]), have implemented only specific algorithms from a research paper \cite{fredrikson2015model} (original library [L.1] and unoriginal libraries [L.30], [L.31]), or observed results that differ from the paper's findings (Our main features are availability, operability, etc.). However, it is important to acknowledge that some original libraries might present usability challenges owing to delays in maintenance by their authors. In these circumstances, the presence of unoriginal libraries can effectively address these concerns.
\subsubsection{\textbf{Choice of Python Versions}}
Additionally, we strongly advise researchers to opt for newer Python versions. This is crucial because some Python libraries may no longer be compatible with older versions. Moreover, older Python versions can introduce complexities during the environment setup phase, such as difficulties in downloading specific Python libraries using standard installation commands.
\subsubsection{\textbf{Common Limitations Across Libraries}}
Each library boasts its unique characteristics and distinctions. Nevertheless, they all share common limitations. For example, some authors may provide overly simplified documentation that proves challenging to comprehend, or certain open-source libraries may offer datasets that are relatively limited in diversity compared to those mentioned in the corresponding research papers. It's worth noting that during our experiments, we observed similarities between certain research papers and libraries. For instance, literature \cite{shokri2017membership} (library [L.1]) and \cite{salem2018ml} (library [L.4]) shared identical core attack methodologies, with the latter introducing more relaxed assumptions to extend the former's work. Furthermore, the Python package FoolBox streamlines available adversarial attack techniques and provides a unified API for various deep learning frameworks.

\section{\textbf{Conclusion}}
This paper delves into an extensive examination of various security attack types pertinent to machine learning as applied in ML-based IoT. We cover a broad spectrum of threats including membership inference attacks, adversarial attacks, reconstruction attacks, property inference attacks, model extraction attacks, and poisoning attacks, all of which present formidable challenges to the integrity, confidentiality, and availability of machine learning-based ML-based IoT frameworks. Our analysis is multi-faceted, encompassing different aspects of the attack landscape, attacker capabilities, and the security properties at risk. We provide an all-encompassing perspective, crucial for understanding and mitigating threats within ML-based IoT environments infused with machine learning. In addition, we have conducted a rigorous evaluation of 65 libraries specifically geared toward machine learning security in ML-based IoT settings. Our work contributes to this ongoing effort, aiming to fortify machine learning applications against malicious actors and unforeseen vulnerabilities in the ever-evolving ML-based IoT landscape. %This enriches the existing body of knowledge and provides practical insights for researchers and practitioners alike. Safeguarding machine learning systems within ML-based IoT networks is imperative. As these systems become increasingly integrated into critical infrastructures and everyday devices, ensuring their security and protecting the vast amounts of data they process is not just a technical necessity but a societal imperative. 

\section{\textbf{Acknowledgement}}
We thank Dr. Yusen Wu for his help and comments. Dr. Chao Liu is supported by the Fujian Province Industry-University-Research Collaboration Project (Project Number: 2023H6034) and the Fujian Province Project for Young and Middle-Aged Researchers (Project Number: JAT220827).

\bibliographystyle{IEEEtran}
\bibliography{ref.bib}
\vspace{-10pt}
\begin{IEEEbiographynophoto}
%[{\includegraphics[width=1in,height=1.25in,clip,keepaspectratio]{Photo/chao.jpg}}]
{Chao Liu}
(M'22) obtained the doctoral degree in Computer Science and Electronic Engineering from the University of Maryland, Baltimore County. He was advised by Dr. Haibin Zhang, Dr. Sisi Duan, and Dr. Alan Sherman. Dr. Chao Liu's research focuses on asynchronous Byzantine fault tolerance (BFT) protocols and consensus algorithms for distributed systems and blockchains. He has published 16 high-level academic papers, including 2 papers in IEEE/IFIP International Conference on Dependable Systems and Networks (DSN), 1 paper in Symposium on Reliable Distributed Systems (SRDS), and 1 paper in the Journal of Parallel and Distributed Computing (JPDC). He also worked with David Chaum, who is the pioneer of anonymous communication, blind signature, group signature, and digital currency. Their work VoteXX appears in the top voting conference E-VOTE-ID 2023. Their team won the grand prize of the innovation track at the 2022 Financial Cryptography Competition. He is a member of the academic committee of CCF YOCSEF Xiamen from 2023 to 2024.
\end{IEEEbiographynophoto}
\vspace{-1cm}
\begin{IEEEbiographynophoto}
%[{\includegraphics[width=1in,height=1.251in,clip,keepaspectratio]{Photo/Boxi_Chen.jpg}}] 
{Boxi Chen} (S'23) conducts research in the field of privacy protection in machine learning. Previously, he completed an internship at the Department of Research at Deep Red Future Technology. Additionally, he has gained valuable experience while working at the Fujian Provincial Key Laboratory of Data Intensive Computing and the Key Laboratory of Intelligent Computing and Information Processing.
\end{IEEEbiographynophoto}
\vspace{-1cm}
\begin{IEEEbiographynophoto}
%[{\includegraphics[width=1in,height=1.251in,clip,keepaspectratio]{Photo/Weishao.png}}]
{Wei Shao} is a Research Scientist at CSIRO Data61, an adjunct fellow at RMIT University, and Visiting Researcher at UC Davis, is currently engrossed in exploring AI security, network security, graph neural networks, spatio-temporal data mining, edge-computing autonomous driving, reinforcement learning, and the Internet of Things.
\end{IEEEbiographynophoto}
\vspace{-0.05cm}
\begin{IEEEbiographynophoto}
%[{\includegraphics[width=1in,height=1.25in,clip,keepaspectratio]{Photo/wenjun.jpg}}]
{Wenjun (Chris) Zhang} received his Ph.D. degree in Mechanical Engineering and Design from the Delft University of Technology, The Netherlands, in 1994. He is currently a full professor of the Department of Mechanical Engineering in the University of Saskatchewan, Canada. He has published over 580 refereed technical publications, among which over 360 papers appear in refereed journals and over 220 papers in refereed conference proceedings, in a broad scope of fields, including design and mechatronics, manufacturing, robotics, sensors/actuators, informatics, and human-machine systems. He holds dozens of patents, including three US patents. His paper received the best journal paper award for 2012 by the IEEE Transactions on Neural Networks. His papers received the top 25 hottest articles by Mechatronics in 2006 \&2005, by Expert Systems with Applications in 2011, by Microelectronics Journal. He has the Google Scholar H-index (66). He was highly cited author by Elsevier(China) from 2015 to 2018(inclusive). He was included in the top 2\% world scientists by Stanford in 2020-2021. He has served as a Senior Editor for the IEEE/ASME Transactions on Mechatronics, and Associate Editor for several reputed journals including IEEE SMC - System, and IEEE System Journal. He is an elected Fellow of the Canadian Academy of Engineering, a fellow of the American Society of Mechanical Engineers, a Fellow of the Society of Manufacturing Engineering, and Senior member of IEEE.
\end{IEEEbiographynophoto}
%\vspace{5pt}
\begin{IEEEbiographynophoto}
%[{\includegraphics[width=1in,height=1.25in,clip,keepaspectratio]{Photo/wong.png}}]
{Kelvin K.L. Wong}
(M’20) obtained a BEng (Hons, 2001) in Mechanical and Production Engineering from Nanyang Technological University, a MAIT (2003) in Applied Information Technology from The University of Sydney, followed by a PhD in Electrical and Electronic Engineering (2009) from The University of Adelaide. Dr. Kelvin Wong combined explainable artificial intelligence and cybernetics in cardiovascular hydrodynamics for the first time. These researches have won international recognition (totaling more than 30 relevant papers). Dr. Kelvin Wong has published more than 100 papers in top journals and internationally recognized academic journals and served as deputy editor and guest editor-in-chief in many well-known academic journals. He developed Deep Red AI cloud processing tool to be used in deep learning-based medical research and data analytics. Dr. Kelvin Wong currently is the professor and doctoral advisor at the Shenzhen Institutes of Advanced Technology, and the professor at the University of Saskatchewan, Canada. He serves as a reviewer for a number of reputable international peer-reviewed journals including IEEE Journal of Biomedical and Health Informatics, IEEE Transactions on Pattern Analysis and Machine Intelligence, IEEE Transactions on Neural Networks and Learning Systems, IEEE Systems, Man, and Cybernetics among others.
\end{IEEEbiographynophoto}

%\vspace{3cm}
\begin{IEEEbiographynophoto}
%[{\includegraphics[width=1in,height=1.25in,clip,keepaspectratio]{Photo/yizhang.png}}]
{Yi Zhang}
(F’15) received the Ph.D. degree in mathematics from the Institute of Mathematics, The Chinese Academy of Science, Beijing, China, in 1994. Currently, he is a professor and the director of the Intelligent Interdisciplinary Research Center at Sichuan University. He was the dean at the College of Computer Science, Sichuan University, Chengdu, China (2008.10-2017.6). He is the co-author of three books: Convergence Analysis of Recurrent Neural Networks (Kluwer Academic Publisher, 2004), Neural Networks: Computational Models and Applications (Springer, 2007), and Subspace Learning of Neural Networks (CRC Press, 2010).
He was an Associate Editor of IEEE Transactions on Neural Networks and Learning Systems (2009–2012), and he is an Associate Editor of IEEE Transactions on Cybernetics (2014–). He is the founding director of Machine Intelligence Laboratory at Sichuan University, and the founding Chair of IEEE Computational Intelligence Society, Chengdu Chapter. Presently, he is the Chair of IEEE Chengdu Section (2015-). His current research interests including Neural Networks and Intelligent Medicine. He is a fellow of IEEE and Russian Academy of Engineering.
\end{IEEEbiographynophoto}

\section{\textbf{Appendix}}
\noindent {[L.1]}\hspace{0.4em}https://github.com/csong27/membership-inference\\
\noindent {[L.2]}\hspace{0.4em}https://github.com/yonsei-cysec/Membership\_Inference\\\_Attack\hfill\\
\noindent {[L.3]}\hspace{0.4em}https://github.com/spring-epfl/mia\\
\noindent {[L.4]}\hspace{0.4em}https://github.com/AhmedSalem2/ML-Leaks\\
\noindent {[L.5]}\hspace{0.4em}https://github.com/GeorgeTzannetos/ml-leaks-pytorch\\
\noindent {[L.6]}\hspace{0.4em}https://github.com/inspire-group/privacy-vs-robustness\\
\noindent {[L.7]}\hspace{0.4em}https://github.com/inspire-group/membership-inference-evaluation\\
\noindent {[L.8]}\hspace{0.4em}https://github.com/jinyuan-jia/MemGuard\\
\noindent {[L.9]}\hspace{0.4em}https://github.com/hyhmia/BlindMI\\
\noindent {[L.10]}\hspace{0.4em}https://github.com/HongshengHu/source-inference-FL\\ 
\noindent {[L.11]}\hspace{0.4em}https://github.com/DennisLiu2022/Membership-Inference-Attacks-by-Exploiting-Loss-Trajectory\hfill\\
\noindent {[L.12]}\hspace{0.4em}https://github.com/shrebox/Privacy-Attacks-in-Machin\\e-Learning\\
\noindent {[L.13]}\hspace{0.4em}https://github.com/privacytrustlab/ml\_privacy\_meter\\
\noindent {[L.14]}\hspace{0.4em}https://github.com/LTS4/DeepFool\\
\noindent {[L.15]}\hspace{0.4em}https://github.com/carlini/nn\_robust\_attacks\\
\noindent {[L.16]}\hspace{0.4em}https://github.com/huanzhang12/ZOO-Attack\\
\noindent {[L.17]}\hspace{0.4em}https://github.com/cg563/simple-blackbox-attack\\
\noindent {[L.18]}\hspace{0.4em}https://github.com/dydjw9/MetaAttack\_ICLR2020\\
\noindent {[L.19]}\hspace{0.4em}https://github.com/xiaosen-wang/TA\\
\noindent {[L.20]}\hspace{0.4em}https://github.com/1Konny/FGSM\\
\noindent {[L.21]}\hspace{0.4em}https://github.com/bethgelab/foolbox\\
\noindent {[L.22]}\hspace{0.4em}https://github.com/EG-Booster/code\\
\noindent {[L.23]}\hspace{0.4em}https://github.com/greentfrapp/boundary-attack\\
\noindent {[L.24]}\hspace{0.4em}https://github.com/takerum/vat\\
\noindent {[L.25]}\hspace{0.4em}https://github.com/Hyperparticle/one-pixel-attack-keras\\
\noindent {[L.26]}\hspace{0.4em}https://github.com/pajola/ZeW\\
\noindent {[L.27]}\hspace{0.4em}https://github.com/labsix/limited-blackbox-attacks\\
\noindent {[L.28]}\hspace{0.4em}https://github.com/Jianbo-Lab/HSJA\\
\noindent {[L.29]}\hspace{0.4em}https://github.com/Trusted-AI/adversarial-robustness-toolbox\\
\noindent {[L.30]}\hspace{0.4em}http://www.cs.cmu.edu/~mfredrik/mi-2016.zip\\
\noindent {[L.31]}\hspace{0.4em}https://github.com/zhangzp9970/MIA\\
\noindent {[L.32]}\hspace{0.4em}https://github.com/zechenghe/Inverse\_Collaborative\_Inf\\erence\\
\noindent {[L.33]}\hspace{0.4em}https://github.com/ege-erdogan/unsplit\\
\noindent {[L.34]}\hspace{0.4em}https://github.com/Jaskiee/GAN-Attack-against-Federated-Deep-Learning\\
\noindent {[L.35]}\hspace{0.4em}https://github.com/m-kahla/Label-Only-Model-Inversion-Attacks-via-Boundary-Repulsion\\
\noindent {[L.36]}\hspace{0.4em}https://github.com/HanGyojin/RLB-MI\\
\noindent {[L.37]}\hspace{0.4em}https://github.com/deepmind/informed\_adversary\_mnist\\\_reconstruction\\
\noindent {[L.38]}\hspace{0.4em}https://github.com/JunyiZhu-AI/R-GAP\\
\noindent {[L.39]}\hspace{0.4em}https://github.com/yziqi/adversarial-model-inversion\\
\noindent {[L.40]}\hspace{0.4em}https://github.com/csong27/ml-model-remember\\
\noindent {[L.41]}\hspace{0.4em}https://github.com/PatrickZH/Improved-Deep-Leakage-from-Gradients\\
\noindent {[L.42]}\hspace{0.4em}https://github.com/shrebox/Privacy-Attacks-in-Machine-Learning\\
\noindent {[L.43]}\hspace{0.4em}https://github.com/Pilladian/attribute-inference-attack\\
\noindent {[L.44]}\hspace{0.4em}https://github.com/csong27/property-inference-collabor\\ative-ml\hfill\\
\noindent {[L.45]}\hspace{0.4em}https://github.com/mmalekzadeh/honest-but-curious-ne\\ts\hfill\\
\noindent [{L.46]}\hspace{0.4em}https://github.com/iamgroot42/FormEstDistRisks\\
\noindent {[L.47]}\hspace{0.4em}https://drive.google.com/file/d/1hu0PhN3pWXe6Lobxi\\PFeYBm8L-vQX2zJ/view\\
\noindent {[L.48]}\hspace{0.4em}https://github.com/epfl-dlab/distribution-inference-risks\\
\noindent {[L.49]}\hspace{0.4em}https://github.com/iamgroot42/dissecting\_dist\_inf\\
\noindent {[L.50]}\hspace{0.4em}https://github.com/matprst/PIA-on-CNN\\
\noindent {[L.51]}\hspace{0.4em}https://github.com/pasquini-dario/SplitNN\_FSHA\\
\noindent {[L.52]}\hspace{0.4em}https://github.com/ftramer/Steal-ML\\
\noindent {[L.53]}\hspace{0.4em}https://github.com/TrustworthyGNN/MEA-GNN\\
\noindent {[L.54]}\hspace{0.4em}https://github.com/ftramer/LM\_Memorization\\
\noindent {[L.55]}\hspace{0.4em}https://github.com/antoniobarbalau/black-box-ripper\\
\noindent {[L.56]}\hspace{0.4em}https://github.com/google-research/cryptanalytic-model\\-extraction\hfill\\
\noindent {[L.57]}\hspace{0.4em}https://github.com/coallaoh/WhitenBlackBox\\
\noindent {[L.58]}\hspace{0.4em}https://github.com/cleverhans-lab/DatasetInferenceForSelfSupervisedModels\\
\noindent {[L.59]}\hspace{0.4em}https://github.com/jeiks/Stealing\_DL\_Models\\
\noindent {[L.60]}\hspace{0.4em}https://github.com/inspire-group/ModelPoisoning\\
\noindent {[L.61]}\hspace{0.4em}https://github.com/git-disl/DataPoisoning\_FL\\
\noindent {[L.62]}\hspace{0.4em}https://github.com/rdz98/FedRecAttack\\
\noindent {[L.63]}\hspace{0.4em}https://github.com/ChangxinTian/RAPU\\
\noindent {[L.64]}\hspace{0.4em}https://github.com/icmpnorequest/MLSec\\
\noindent {[L.65]}\hspace{0.4em}https://github.com/suyeecav/model-targeted-poisoning\\


\end{document}